\documentclass[]{pasj01}
\usepackage{graphics}
\usepackage{color}
\begin{document}
\SetRunningHead{N. Isobe et al.}
{X-ray and IR diagnostics of nearby active galactic nuclei}
\Received{2016/06/20}
\Accepted{2016/08/23}

%%%%%%%%%
% Title %
%%%%%%%%%
\title{X-ray and % mid-to-far 
infrared diagnostics of nearby active galactic nuclei with MAXI and AKARI.}

%%%%%%%%%%%
% authors %   
%%%%%%%%%%%
\author{%
Naoki       \textsc{Isobe}     \altaffilmark{1,2},
Taiki       \textsc{Kawamuro}  \altaffilmark{3},
Shinki      \textsc{Oyabu}     \altaffilmark{4},
Takao       \textsc{Nakagawa}  \altaffilmark{2},
Shunsuke    \textsc{Baba}      \altaffilmark{2,5},
Kenichi    \textsc{Yano}      \altaffilmark{2,5},
Yoshihiro   \textsc{Ueda}      \altaffilmark{3},
\&
Yoshiki     \textsc{Toba}      \altaffilmark{6,7}
}
\altaffiltext{1}{
	School of Science, Tokyo Institute of Technology, 
	2-12-1 Ookayama, Meguro, Tokyo 152-8551, Japan}
\email{n-isobe@hp.phys.titech.ac.jp}
\altaffiltext{2}{
  Institute of Space and Astronautical Science (ISAS), 
  Japan Aerospace Exploration Agency (JAXA), 
  3-1-1 Yoshinodai, Chuo-ku, Sagamihara, Kanagawa 252-5210, Japan}
\altaffiltext{3}{
  Department of Astronomy, Kyoto University, Oiwake-cho, 
  Sakyo-ku, Kyoto 606-8502, Japan}
\altaffiltext{4}{
   % Division of Particle and Astronphysical Science, 
  Graduate School of Science, Nagoya University, Furo-cho,
  Chikusa-ku, Nagoya, Aichi 464-8602, Japan}
\altaffiltext{5}{
   Department of Physics, The University of Tokyo, 
   7-3-1, Hongo, Bunkyo-ku, Tokyo, 133-0033, Japan
}
\altaffiltext{6}{
  Research Center for Space and Cosmic Evolution,
  Ehime University, 
  2-5 Bunkyo-cho, Matsuyama, Ehime 790-8577, Japan}
\altaffiltext{7}{
  Academia Sinica Institute of Astronomy and Astrophysics, 
  PO Box 23-141, Taipei 10617, Taiwan
 }

\maketitle

\begin{abstract}
Nearby active galactic nuclei were diagnosed 
in the X-ray and mid-to-far infrared wavelengths,
with Monitor of All-sky X-ray Image (MAXI) 
and the Japanese infrared observatory AKARI, respectively.
Among the X-ray sources 
listed in the second release of the MAXI all-sky X-ray source catalog,
100 ones are currently identified 
as a non-blazar-type active galactic nucleus.
These include 95 Seyfert galaxies and 5 quasars, 
and they are composed of 73 type-1 and 27 type-2 objects.
The AKARI all-sky survey point source catalog was searched 
for their mid- and far-infrared counterparts at $9$, $18$, and $90$ $\mu$m.
As a result, 69 Seyfert galaxies in the MAXI catalog 
(48 type-1 and 21 type-2 ones) were found to be detected with AKARI.
The X-ray ($3$--$4$ keV and $4$--$10$ keV) and infrared luminosities 
of these objects were investigated, together with their color information.
Adopting the canonical photon index, $\Gamma = 1.9$, 
of the intrinsic X-ray spectrum of the Seyfert galaxies,
the X-ray hardness ratio 
between the $3$--$4$ and $4$--$10$ keV ranges derived with MAXI 
was roughly converted into the absorption column density.
After the X-ray luminosity was corrected for absorption 
from the estimated column density,
the well-known X-ray-to-infrared luminosity correlation 
was confirmed at least in the Compton-thin regime.
In contrast, NGC 1365, only one Compton-thick object in the MAXI catalog,
was found to deviate from the correlation 
toward a significantly lower X-ray luminosity by nearly an order of magnitude.
It was verified that 
the relation between the X-ray hardness below 10 keV and X-ray-to-infrared color 
acts as an effective tool to pick up Compton-thick objects.
The difference in the infrared colors 
between the type-1 and type-2 Seyfert galaxies 
and its physical implication on the classification and unification of active galactic nuclei
were briefly discussed. 
\end{abstract}

%%%%%%%%%%%%%
% Section 1 %
%%%%%%%%%%%%%
\section{Introduction}  %=============================================
\label{sec:intro}
The unified picture of active galactic nuclei 
(e.g., \cite{AGN_Unify_1,AGN_Unify_2})
invokes a supermassive black hole accompanied by a mass accreting disk 
as their central engine.
It is widely believed that they are surrounded 
by a parsec-scale torus filled with dust and gas clouds. 
However, the detailed physical condition and spatial configuration
of these essential ingredients in the active galactic nuclei
still remain one of the important issues 
yet to be solved in the modern astrophysics.

The active galactic nuclei are widely known as an X-ray emitter.
Except for jet-dominated sources including blazars and BL Lacertae objects, 
their X-ray emission is thought to originate 
in a hot corona above the accretion disk,
where disk photons are Comptonized (e.g., \cite{Comptonization}).
After the nuclear radiation is absorbed by the dust torus,
it is re-emitted into the mid-to-far infrared (IR) wavelength.
As a result, the active galactic nuclei,
without any significant jet contamination, 
typically exhibit a strong IR bump 
in their spectral energy distribution around $1$ -- $100$ $\mu$m 
(e.g., \cite{AGN_IR,AGN_Unify_1,QSO_atras,AGN_IRbump}). 
These properties make a combination of X-ray and IR observations 
an ideal probe for the central region of the active galactic nuclei.
Actually, a number of recent studies indicate 
a linear correlation between the absorption-corrected 
(or -unaffected) X-ray luminosities 
and the observed IR ones among nearby active galactic nuclei 
with a moderate absorbing Hydrogen column density of 
$N_{\rm H} \lesssim 10^{24}$ cm$^{-2}$,
regardless of their optical classification 
\citep{LIR-LX_AGN,Swift/BAT-AKARI_AGN1,Swift/BAT-AKARI_AGN2}.
Such a correlation is supposed to 
prefer a so-called clumpy torus geometry (e.g., \cite{clumpy_torus}),
instead of a simple torus model 
with a smooth and homogeneous dust distribution (e.g., \cite{smooth_torus}),
because the latter model requests 
a deficit in the observed IR luminosity from obscured objects 
due to self-extinction within the dust torus.

%---------%
% Table 1 %
%---------%
%---------%
% Table 1 %
%---------%
\begin{table*}[p]
\caption{X-ray and IR properties of the active galactic nuclei, 
         detected both with MAXI and AKARI\footnotemark[$*$].}
\label{tab:catalog}
\begin{center}
{\tiny
\begin{tabular}[c]{llllllllll}
\hline %==================================================
(1)   & (2)         & (3)       & (4)        & (5)  & (6)       &  (7)        & (8)        & (9) & (10)      \\
2MAXI & Object Name & $F_{\rm S}$ & $F_{\rm H}$ & $HR$ & $f_{\rm 9}$ & $f_{\rm 18}$ & $f_{\rm 90}$ & $z$ & type \\ 
\hline 
 J0044$-$238 &                 NGC 235A & $ 0.6 \pm 0.3$   &  $  7.8 \pm 1.1$   &  $ 0.64 \pm  0.18$   &                 ---   &  $   295 \pm    32$   &                 ---   &  $  0.0222 $  &      Sy1 \\
 J0048$+$320 &                  Mrk 348 & $ 2.8 \pm 0.4$   &  $ 21.4 \pm 1.3$   &  $ 0.42 \pm  0.06$   &                 ---   &  $   593 \pm    43$   &  $   736 \pm    60$   &  $   0.015 $  &      Sy2 \\
                                                                                                                                                                                                %Sy2/FSRQ \\
 J0124$-$349 &                 NGC 526A & $ 5.4 \pm 0.4$   &  $ 22.5 \pm 1.2$   &  $ 0.16 \pm  0.04$   &  $   141 \pm    14$   &  $   292 \pm    26$   &                 ---   &  $  0.0191 $  &    Sy1.5 \\
 J0124$-$588 &             Fairall 0009 & $ 4.3 \pm 0.3$   &  $ 17.9 \pm   1$   &  $ 0.15 \pm  0.04$   &  $   229 \pm    20$   &  $   440 \pm    24$   &                 ---   &  $   0.047 $  &      Sy1 \\
 J0228$+$314 &                  NGC 931 & $ 4.3 \pm 0.4$   &  $   18 \pm 1.5$   &  $ 0.15 \pm  0.06$   &  $   349 \pm    12$   &  $   763 \pm    48$   &  $  2430 \pm    64$   &  $  0.0167 $  &    Sy1.5 \\
 J0233$+$324 &                  NGC 973 & $ 3.9 \pm 0.4$   &  $ 16.6 \pm 1.5$   &  $ 0.17 \pm  0.07$   &                 ---   &  $    93 \pm    77$   &  $  1704 \pm    69$   &  $  0.0162 $  &      Sy2 \\
 J0334$-$360 &                 NGC 1365 & $ 3.6 \pm 0.4$   &  $ 12.9 \pm 1.2$   &  $ 0.07 \pm  0.07$   &  $  2234 \pm    38$   &  $  5364 \pm    42$   &  $ 80384 \pm 13546$   &  $  0.0055 $  &    Sy1.8 \\
 J0342$-$214 &           ESO 548$-$G081 & $ 3.8 \pm 0.3$   &  $ 14.6 \pm 1.1$   &  $ 0.11 \pm  0.06$   &                 ---   &                 ---   &  $   968 \pm    61$   &  $  0.0145 $  &      Sy1 \\
 J0424$-$571 &            1H 0419$-$577 & $ 2.6 \pm 0.3$   &  $ 13.4 \pm 1.1$   &  $ 0.25 \pm  0.06$   &                 ---   &  $   106 \pm    21$   &                 ---   &  $   0.104 $  &    Sy1.5 \\
 J0433$+$054 &                   3C 120 & $ 8.1 \pm 0.4$   &  $ 25.4 \pm 1.3$   &  $ 0.01 \pm  0.04$   &  $   203 \pm    20$   &  $   497 \pm    68$   &  $  1468 \pm    85$   &  $   0.033 $  &      Sy1 \\
%---10 -----------------------------------------------
 J0437$-$106 &         MCG $-$02-12-050 & $ 3.2 \pm 0.3$   &  $ 12.5 \pm 1.2$   &  $ 0.12 \pm  0.07$   &                 ---   &                 ---   &  $   682 \pm   111$   &  $  0.0364 $  &    Sy1.2 \\
 J0443$+$288 &                UGC 03142 & $ 1.7 \pm 0.4$   &  $ 11.8 \pm 1.2$   &  $ 0.38 \pm   0.1$   &                 ---   &                 ---   &  $  1581 \pm    79$   &  $  0.0217 $  &      Sy1 \\
 J0453$-$752 &            ESO 033-G 002 & $ 4.1 \pm 0.3$   &  $ 13.4 \pm 0.7$   &  $ 0.03 \pm  0.04$   &  $   163 \pm     7$   &  $   387 \pm    15$   &  $   646 \pm   120$   &  $  0.0181 $  &      Sy2 \\ 
 J0516$-$001 &                   Ark120 & $ 6.3 \pm 0.4$   &  $ 20.8 \pm 1.1$   &  $ 0.04 \pm  0.04$   &  $   252 \pm    18$   &  $   253 \pm    33$   &                 ---   &  $  0.0323 $  &      Sy1 \\
 J0516$-$102 &            MCG-02-14-009 & $ 2.2 \pm 0.3$   &  $  8.7 \pm   1$   &  $ 0.13 \pm  0.09$   &  $    90 \pm    21$   &                 ---   &  $   528 \pm    56$   &  $  0.0285 $  &      Sy1 \\
 J0518$-$325 &               ESO 362-18 & $ 2.7 \pm 0.4$   &  $  9.4 \pm 1.1$   &  $ 0.07 \pm  0.09$   &  $   166 \pm    31$   &  $   366 \pm    36$   &  $  1277 \pm    88$   &  $  0.0125 $  &    Sy1.5 \\
 J0523$-$363 &            PKS 0521$-$36 & $ 3.5 \pm 0.4$   &  $ 12.7 \pm 1.2$   &  $ 0.08 \pm  0.07$   &  $  96.6 \pm   0.1$   &  $   216 \pm    20$   &                 ---   &  $  0.0553 $  &    BL Lac \\
 J0552$-$073 &                 NGC 2110 & $13.3 \pm 0.4$   &  $ 71.1 \pm 1.3$   &  $ 0.27 \pm  0.02$   &  $   300 \pm    19$   &  $   566 \pm    30$   &  $  4594 \pm    47$   &  $  0.0078 $  &      Sy2 \\
 J0554$+$464 &         MCG $+$08-11-011 & $ 9.2 \pm 0.4$   &  $ 33.7 \pm 1.3$   &  $ 0.09 \pm  0.03$   &  $   340 \pm    18$   &  $  1283 \pm    35$   &  $  2377 \pm    62$   &  $  0.0205 $  &    Sy1.5 \\
 J0615$+$712 &                    Mrk 3 & $ 0.4 \pm 0.3$   &  $  7.2 \pm 0.9$   &  $ 0.69 \pm  0.17$   &  $   322 \pm     5$   &  $  1892 \pm    49$   &  $  2939 \pm   282$   &  $  0.0135 $  &      Sy2 \\
%---20 -----------------------------------------------
 J0640$-$257 &            ESO 490-IG026 & $ 2.6 \pm 0.3$   &  $  7.8 \pm   1$   &  $-0.01 \pm  0.09$   &                 ---   &                 ---   &  $  1409 \pm    41$   &  $  0.0248 $  &    Sy1.2 \\
 J0740$+$497 &                   Mrk 79 & $ 2.8 \pm 0.4$   &  $  8.9 \pm 1.2$   &  $ 0.03 \pm   0.1$   &  $   276 \pm     6$   &  $   611 \pm    38$   &  $  1358 \pm    64$   &  $  0.0222 $  &    Sy1.2 \\
 J0804$+$052 &                 Mrk 1210 & $ 0.9 \pm 0.4$   &  $  9.7 \pm 1.2$   &  $ 0.55 \pm  0.14$   &  $   274 \pm    19$   &  $  1310 \pm     6$   &  $  1196 \pm    61$   &  $  0.0135 $  &      Sy2 \\
 J0808$+$757 &            PG 0804$+$761 & $   3 \pm 0.3$   &  $  8.5 \pm 0.9$   &  $-0.03 \pm  0.07$   &  $   118 \pm    10$   &  $   186 \pm    15$   &                 ---   &  $     0.1 $  &      Sy1 \\
 J0920$-$079 &         MCG $-$01-24-012 & $ 3.6 \pm 0.4$   &  $   17 \pm 1.2$   &  $ 0.21 \pm  0.06$   &                 ---   &  $   263 \pm    44$   &                 ---   &  $  0.0196 $  &      Sy2 \\
 J0923$+$228 &         MCG $+$04-22-042 & $ 2.3 \pm 0.4$   &  $ 10.3 \pm 1.1$   &  $  0.2 \pm  0.09$   &  $    78 \pm    15$   &  $   178 \pm    47$   &                 ---   &  $  0.0323 $  &    Sy1.2 \\
 J0945$-$140 &                 NGC 2992 & $ 1.3 \pm 0.3$   &  $  9.2 \pm   1$   &  $ 0.39 \pm  0.11$   &  $   299 \pm    49$   &  $   826 \pm    54$   &  $  9220 \pm   194$   &  $  0.0077 $  &      Sy2 \\
 J0947$-$308 &         MCG $-$05-23-016 & $18.9 \pm 0.4$   &  $ 82.9 \pm 1.3$   &  $ 0.18 \pm  0.01$   &  $   384 \pm    14$   &  $  1391 \pm    21$   &  $  1277 \pm    84$   &  $  0.0085 $  &      Sy2 \\
 J1000$-$313 & 2MASX J09594263$-$3112581& $ 3.1 \pm 0.4$   &  $ 14.3 \pm 1.1$   &  $ 0.19 \pm  0.07$   &  $    85 \pm     1$   &  $   234 \pm     5$   &                 ---   &  $   0.037 $  &      Sy1 \\
 J1023$+$199 &                 NGC 3227 & $ 5.8 \pm 0.4$   &  $ 23.2 \pm 1.2$   &  $ 0.14 \pm  0.04$   &  $   444 \pm    71$   &  $  1128 \pm    44$   &  $ 10596 \pm   535$   &  $  0.0039 $  &    Sy1.5 \\
%---30 -----------------------------------------------
 J1031$-$143 & 2MASSi J1031543$-$141651 & $ 2.4 \pm 0.3$   &  $ 11.7 \pm 1.1$   &  $ 0.22 \pm  0.08$   &  $    94 \pm     1$   &                 ---   &                 ---   &  $   0.086 $  &      Sy1 \\
 J1105$+$724 &                 NGC 3516 & $ 4.1 \pm 0.3$   &  $ 24.8 \pm   1$   &  $ 0.33 \pm  0.04$   &  $   262 \pm    20$   &  $   651 \pm    16$   &  $  1317 \pm    83$   &  $  0.0088 $  &    Sy1.5 \\
 J1139$-$376 &                 NGC 3783 & $ 8.4 \pm 0.4$   &  $ 39.7 \pm 1.3$   &  $ 0.21 \pm  0.03$   &  $   502 \pm    10$   &  $  1530 \pm    41$   &  $  2716 \pm   108$   &  $  0.0097 $  &      Sy1 \\
 J1203$+$445 &                 NGC 4051 & $ 5.2 \pm 0.4$   &  $ 18.8 \pm 1.2$   &  $ 0.08 \pm  0.05$   &  $   346 \pm    30$   &  $   885 \pm    42$   &  $  4557 \pm   254$   &  $  0.0023 $  &    Sy1.5 \\
 J1210$+$394 &                 NGC 4151 & $15.1 \pm 0.5$   &  $122.6 \pm 1.3$   &  $ 0.45 \pm  0.01$   &  $  1032 \pm    19$   &  $  3629 \pm    72$   &  $  4594 \pm   126$   &  $  0.0033 $  &    Sy1.5 \\
 J1217$+$073 &                 NGC 4235 & $ 2.3 \pm 0.4$   &  $ 11.3 \pm 1.2$   &  $ 0.24 \pm  0.09$   &                 ---   &                 ---   &  $   401 \pm    50$   &  $   0.008 $  &      Sy1 \\
 J1325$-$429 &              Centaurus A & $34.1 \pm 0.5$   &  $300.2 \pm 1.8$   &  $ 0.48 \pm  0.01$   &  $ 10191 \pm  2305$   &  $ 13148 \pm  1049$   &  $102187 \pm 12824$   &  $  0.0018 $  &      Sy2 \\
 J1335$-$341 &         MCG $-$06-30-015 & $ 8.7 \pm 0.4$   &  $ 36.2 \pm 1.2$   &  $ 0.15 \pm  0.03$   &  $   280 \pm    32$   &  $   591 \pm    11$   &  $  1035 \pm   119$   &  $  0.0077 $  &    Sy1.2 \\
 J1338$+$046 &                 NGC 5252 & $ 3.5 \pm 0.4$   &  $ 22.4 \pm 1.2$   &  $ 0.36 \pm  0.05$   &                 ---   &                 ---   &  $   416 \pm   120$   &  $   0.023 $  &    Sy1.9 \\
 J1349$-$302 &                 IC 4329A & $18.7 \pm 0.4$   &  $ 75.7 \pm 1.2$   &  $ 0.14 \pm  0.01$   &  $   769 \pm    12$   &  $  1790 \pm    34$   &  $  1785 \pm   210$   &  $   0.016 $  &    Sy1.2 \\
%---40 -----------------------------------------------
 J1413$-$030 &                 NGC 5506 & $12.5 \pm 0.4$   &  $ 60.3 \pm 1.2$   &  $ 0.22 \pm  0.02$   &  $   823 \pm    27$   &  $  2240 \pm    69$   &  $  8413 \pm   302$   &  $  0.0062 $  &    Sy1.9 \\
 J1417$+$253 &                 NGC 5548 & $ 8.1 \pm 0.4$   &  $ 28.8 \pm 1.5$   &  $ 0.08 \pm  0.03$   &  $   157 \pm     5$   &  $   409 \pm    40$   &  $  1073 \pm   237$   &  $  0.0172 $  &    Sy1.5 \\
 J1419$-$265 &             ESO 511-G030 & $ 1.9 \pm 0.3$   &  $ 10.1 \pm 1.1$   &  $ 0.28 \pm   0.1$   &                 ---   &                 ---   &  $   847 \pm   150$   &  $  0.0224 $  &      Sy1 \\
 J1423$+$250 &                 NGC 5610 & $ 2.1 \pm 0.4$   &  $ 11.2 \pm 1.5$   &  $ 0.27 \pm   0.1$   &  $   167 \pm    19$   &  $   371 \pm    41$   &  $  5795 \pm   205$   &  $  0.0169 $  &      Sy2 \\
 J1437$+$588 &                  Mrk 817 & $ 2.7 \pm 0.3$   &  $  9.6 \pm   1$   &  $ 0.07 \pm  0.08$   &  $   188 \pm    10$   &  $   669 \pm    27$   &  $  1575 \pm    60$   &  $  0.0314 $  &    Sy1.5 \\
 J1512$-$213 & 2MASX J15115979$-$2119015& $ 2.5 \pm 0.3$   &  $ 10.3 \pm 1.1$   &  $ 0.14 \pm  0.08$   &  $    95 \pm     8$   &                 ---   &  $  1591 \pm    89$   &  $  0.0446 $  &      Sy1 \\
                                                                                                                                                                                                   % Sy1/NL \\
 J1513$+$423 &                 NGC 5899 & $   2 \pm 0.4$   &  $  8.9 \pm 1.2$   &  $ 0.19 \pm  0.12$   &                 ---   &                 ---   &  $  4683 \pm   184$   &  $  0.0086 $  &      Sy2 \\
 J1535$+$581 &                  Mrk 290 & $ 2.3 \pm 0.3$   &  $    8 \pm   1$   &  $ 0.06 \pm  0.09$   &                 ---   &  $   151 \pm    13$   &                 ---   &  $  0.0296 $  &      Sy1 \\
 J1548$-$136 &                 NGC 5995 & $ 2.6 \pm 0.3$   &  $ 13.9 \pm   1$   &  $ 0.27 \pm  0.07$   &  $   325 \pm    19$   &  $   671 \pm     4$   &  $  4580 \pm   329$   &  $  0.0252 $  &      Sy2 \\
 J1555$-$793 &            PKS 1549$-$79 & $ 2.1 \pm 0.3$   &  $    7 \pm 0.8$   &  $ 0.04 \pm  0.08$   &  $    82 \pm     4$   &  $   328 \pm    45$   &  $   925 \pm    95$   &  $  0.1501 $  &      Sy1 \\
%---50 -----------------------------------------------
 J1614$+$660 &                  Mrk 876 & $ 2.4 \pm 0.3$   &  $  7.5 \pm 0.8$   &  $ 0.01 \pm  0.08$   &  $    82 \pm     5$   &  $   178 \pm    19$   &  $   541 \pm    39$   &  $   0.129 $  &      Sy1 \\
 J1717$-$628 &                 NGC 6300 & $ 0.4 \pm 0.3$   &  $ 14.2 \pm   1$   &  $ 0.85 \pm   0.1$   &  $   277 \pm    24$   &  $  1336 \pm    97$   &  $ 14928 \pm  1066$   &  $  0.0037 $  &      Sy2 \\
 J1834$+$327 &                   3C 382 & $ 7.3 \pm 0.4$   &  $ 29.1 \pm 1.3$   &  $ 0.13 \pm  0.04$   &  $   120 \pm    12$   &                 ---   &                 ---   &  $  0.0579 $  &      Sy1 \\
 J1836$-$594 &             Fairall 0049 & $   4 \pm 0.3$   &  $ 14.8 \pm 0.9$   &  $ 0.09 \pm  0.05$   &  $   411 \pm    20$   &  $   920 \pm    59$   &  $  2619 \pm   180$   &  $  0.0202 $  &      Sy2 \\
 J1838$-$654 &              ESO 103-035 & $ 2.5 \pm 0.3$   &  $ 25.4 \pm 0.9$   &  $ 0.55 \pm  0.04$   &  $   300 \pm    25$   &  $  1446 \pm    12$   &  $  1227 \pm    77$   &  $  0.0133 $  &      Sy2 \\
 J1842$+$797 &                 3C 390.3 & $ 7.7 \pm 0.3$   &  $ 30.8 \pm 0.9$   &  $ 0.13 \pm  0.03$   &  $    90 \pm    10$   &  $   242 \pm    17$   &                 ---   &  $  0.0561 $  &      Sy1 \\
 J1845$-$626 &             Fairall 0051 & $ 1.5 \pm 0.3$   &  $  7.1 \pm 0.8$   &  $ 0.21 \pm  0.11$   &  $   301 \pm     7$   &  $   697 \pm    62$   &  $  1705 \pm   180$   &  $  0.0142 $  &      Sy1 \\
 J1921$-$587 &             ESO 141-G055 & $   8 \pm 0.3$   &  $ 26.2 \pm   1$   &  $ 0.04 \pm  0.03$   &  $   150 \pm     5$   &  $   233 \pm    38$   &                 ---   &  $   0.036 $  &      Sy1 \\
 J1937$-$060 & 2MASX J19373299$-$0613046& $ 5.8 \pm 0.4$   &  $ 19.1 \pm 1.1$   &  $ 0.04 \pm  0.04$   &                 ---   &  $   718 \pm    28$   &  $  3709 \pm   127$   &  $  0.0103 $  &    Sy1.5 \\
 J2009$-$611 &                 NGC 6860 & $ 3.4 \pm 0.3$   &  $ 12.1 \pm 0.9$   &  $ 0.08 \pm  0.06$   &  $   155 \pm    13$   &  $   357 \pm    61$   &  $  1369 \pm    73$   &  $  0.0149 $  &      Sy1 \\
%---60 -----------------------------------------------
 J2042$+$751 &              4C $+$74.26 & $ 5.2 \pm 0.3$   &  $ 21.4 \pm 0.9$   &  $ 0.15 \pm  0.04$   &  $   147 \pm     6$   &  $   175 \pm     9$   &                 ---   &  $   0.104 $  &      Sy1 \\
 J2044$-$106 &                  Mrk 509 & $11.6 \pm 0.4$   &  $ 42.5 \pm 1.2$   &  $ 0.09 \pm  0.02$   &  $   247 \pm    18$   &  $   499 \pm    17$   &                 ---   &  $  0.0344 $  &    Sy1.2 \\
 J2115$+$821 & 2MASX J21140128$+$8204483& $ 2.7 \pm 0.3$   &  $  9.6 \pm 0.9$   &  $ 0.07 \pm  0.07$   &  $    71 \pm     5$   &  $   105 \pm    32$   &                 ---   &  $   0.084 $  &      Sy1 \\
 J2136$-$624 &  1RXS J213623.1$-$622400 & $ 4.1 \pm 0.3$   &  $ 18.4 \pm 0.9$   &  $  0.2 \pm  0.04$   &                 ---   &  $   142 \pm    11$   &                 ---   &  $  0.0588 $  &      Sy1 \\
 J2200$+$105 &                  Mrk 520 & $ 1.3 \pm 0.4$   &  $  9.6 \pm 1.2$   &  $ 0.41 \pm  0.12$   &  $   146 \pm    12$   &  $   320 \pm    11$   &  $  5040 \pm   100$   &  $  0.0266 $  &    Sy1.9 \\
 J2201$-$317 &                 NGC 7172 & $ 2.2 \pm 0.4$   &  $ 25.2 \pm 1.2$   &  $ 0.57 \pm  0.05$   &  $   316 \pm    17$   &  $   424 \pm    44$   &  $  8087 \pm   218$   &  $  0.0087 $  &      Sy2 \\
 J2238$-$126 &                  Mrk 915 & $   2 \pm 0.3$   &  $  8.4 \pm   1$   &  $ 0.17 \pm   0.1$   &                 ---   &  $   482 \pm   284$   &                 ---   &  $  0.0241 $  &      Sy1 \\
 J2303$+$086 &                 NGC 7469 & $ 6.6 \pm 0.4$   &  $ 21.7 \pm 1.2$   &  $ 0.03 \pm  0.04$   &  $   767 \pm    17$   &  $  2692 \pm    60$   &  $ 27694 \pm  1738$   &  $  0.0163 $  &    Sy1.2 \\
 J2304$-$085 &                  Mrk 926 & $10.5 \pm 0.4$   &  $   38 \pm 1.2$   &  $ 0.09 \pm  0.02$   &  $    60 \pm     3$   &  $   214 \pm    36$   &  $   647 \pm   119$   &  $  0.0469 $  &    Sy1.5 \\
 J2318$+$001 &                 NGC 7603 & $ 3.6 \pm 0.4$   &  $ 13.4 \pm 1.2$   &  $  0.1 \pm  0.07$   &  $   295 \pm    11$   &  $   321 \pm    12$   &  $  1340 \pm   104$   &  $  0.0295 $  &    Sy1.5 \\
%----  70 sources -------------------------------------------------
\hline       %--------------------------------------------------
\multicolumn{10}{l}{\footnotemark[$*$] Remarks:} \\
\multicolumn{10}{l}{(1) Source name in the 2MAXI catalog. }\\
\multicolumn{10}{l}{(2) Name of the optical counterpart. } \\
\multicolumn{10}{l}{(3), (4) Soft ($3$ -- $4$  keV) and hard ($4$ -- $10$ keV) band X-ray flux 
                         in the unit of $10^{-12}$ ergs cm$^{-2}$ s$^{-1}$ \citep{2MAXI}}\\
\multicolumn{10}{l}{(5) X-ray Hardness ratio, defined as $HR=(F'_{\rm H} - F'_{\rm S})/(F'_{\rm H} + F'_{\rm S})$, 
                        where $F'_{\rm H}$ and $F'_{\rm S}$ is the hard and soft band fluxes in the Crab unit;} \\
\multicolumn{10}{l}{~~~i.e., $F'_{\rm H} = F_{\rm H}/ 1.21 \times 10^{-8} {\rm ~ergs~cm}^{-2} {\rm ~s}^{-1}$ 
                        and   $F'_{\rm S} = F_{\rm S}/ 3.98 \times 10^{-9} {\rm ~ergs~cm}^{-2} {\rm ~s}^{-1}$ 
                        (see \cite{2MAXI})}. \\
\multicolumn{10}{l}{(6) -- (8) IR flux density at $9~\mu$m, $18~\mu$m and $90~\mu$m in mJy of the AKARI counterpart 
                        \citep{IRC_Catalog,FIS_Catalog}} \\
\multicolumn{10}{l}{(9) Object redshift. } \\
\multicolumn{10}{l}{(10) Optical Seyfert type (\cite{2MAXI}; and reference therein). } \\
\end{tabular}
}
\end{center}
\end{table*}

Owing to tremendous progress in multi-wavelength observations,
active galactic nuclei that are deeply enshrouded by the dust 
(e.g., \cite{new-type_AGN}) have gradually been uncovered. 
Among such heavily obscured active galactic nuclei, 
those with $N_{\rm H} > 1.5 \times 10^{24}$ cm$^{-2}$ 
are widely called as Compton-thick objects.
In spite of their astrophysical importance for various reasons,
such as the origin of the X-ray background radiation \citep{XRB_Gilli,XRB_Ueda},
the Compton-thick sources are rather elusive in optical observations
because of severe dust extinction.
Even hard X-ray surveys performed with the Burst Alert Telescope (BAT) 
onboard the Swift observatory is inferred to have eventually undercounted 
the Compton-thick objects by a factor of $\sim4$ \citep{Swift/BAT_CTAGN},
although hard X-ray photons above 10 keV are 
expected to exhibit a high penetrating power. 
In contrast, it is recently proposed that 
a combination of X-ray and IR color information
is very useful to select the heavily obscured population
\citep{CT-AGN_XMM-IRAS,CT-AGN_XMM-AKARI}. 

Previous unbiased X-ray and IR diagnostics on nearby active galactic nuclei
are usually based on the hard X-ray survey above 10 keV 
\citep{Swift/BAT-AKARI_AGN1,Swift/BAT-AKARI_AGN2}. 
In contrast, the $2$--$10$ keV X-ray spectral information,
obtained with recent X-ray telescopes,
such as Suzaku \citep{Suzaku}, Chandra, and XMM-Newton, 
is one of the most standard and powerful tools
to investigate the properties of active galaxies.
It is important to compare results from the detailed spectral analysis  
with those from all-sky X-ray surveys 
in the same energy range, namely below 10 keV. 
However, the $2$--$10$ keV samples of active galactic nuclei
adopted for the X-ray-to-IR studies were frequently  
constructed from a restricted sky field (e.g., \cite{LIR-LX_AGN}). 
Therefore, unbiased all-sky survey data below 10 keV with a high sensitivity 
has been strongly requested. 

In the present study, we overcome such situation, 
by making use of the X-ray source catalog  
with the Monitor of All-sky X-ray Image (MAXI; \cite{MAXI})
and the IR one with the Japanese space IR observatory AKARI \citep{AKARI}.
MAXI has continuously monitored all the X-ray sky 
below 10 keV with the sensitivity higher than 
any other previous all-sky X-ray survey missions. 
In the second release of the MAXI all-sky X-ray source catalog
(hereafter the 2MAXI catalog; \cite{2MAXI}),
more than 100 active galactic nuclei, 
including Seyfert galaxies, quasars and blazars, are listed.
Considering the size of the dust torus 
(typically a parsec scale, corresponding to the light-crossing time of a few years),
the X-ray flux averaged over $\sim3$ years, 
which is provide by the 2MAXI catalog, 
is valuable rather than snap-shot data obtained with the pointing X-ray satellites.
The hardness ratio within the MAXI energy range is regarded 
as a good indicator of the absorption column density of the dust torus. 
The ability of MAXI for studies of active galactic nuclei was demonstrated 
by several authors (e.g., \cite{Mrk421_MAXI_1,Mrk421_MAXI_2,CenA_MAXI}).

The IR characteristics of the active galactic nuclei,
picked up from the 2MAXI catalog, 
were examined with the AKARI all-sky survey Point Source Catalog (AKARI/PSC).
With the two IR instruments, the InfraRed Camera (IRC; \cite{IRC}) 
and the Far-Infrared Surveyor (FIS; \cite{FIS}),
the AKARI/PSC widely covers the mid-to-far IR sources 
in the $9$--$160$ $\mu$m range, 
where the dust torus is expected to radiate a significant fraction of its energy. 
Therefore, a combined use of the 2MAXI catalog and AKARI/PSC
helps us to investigate the vicinity of the active galactic nuclei,
and to reveal the nature of the dust torus.

%-----------------------------------------------------%
% Figure 1: 
% Angular separation between the 2MAXI sources  %
% and their AKARI counterpart                   %
%-----------------------------------------------------%
\begin{figure}[t]
\begin{center}
\FigureFile(75mm,75mm){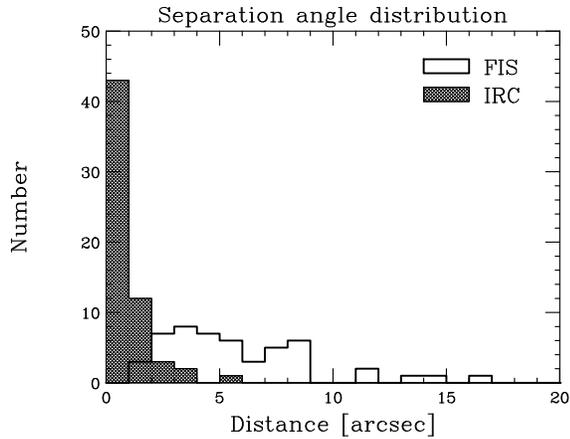}
\end{center}
\caption{Distribution of the separation angle 
between the optical position of the  Seyfert galaxies 
in the 2MAXI catalog and their AKARI counterparts. 
The hatched and thick histograms indicate 
the distributions of the IRC and FIS counterparts, respectively.}
\label{fig:dist_AngSep}
\end{figure}
%-------------------------------------------%
% Figure 2: Redshift and flux distributions %
%-------------------------------------------%
\begin{figure*}[t]
\begin{center}
\FigureFile(75mm,75mm){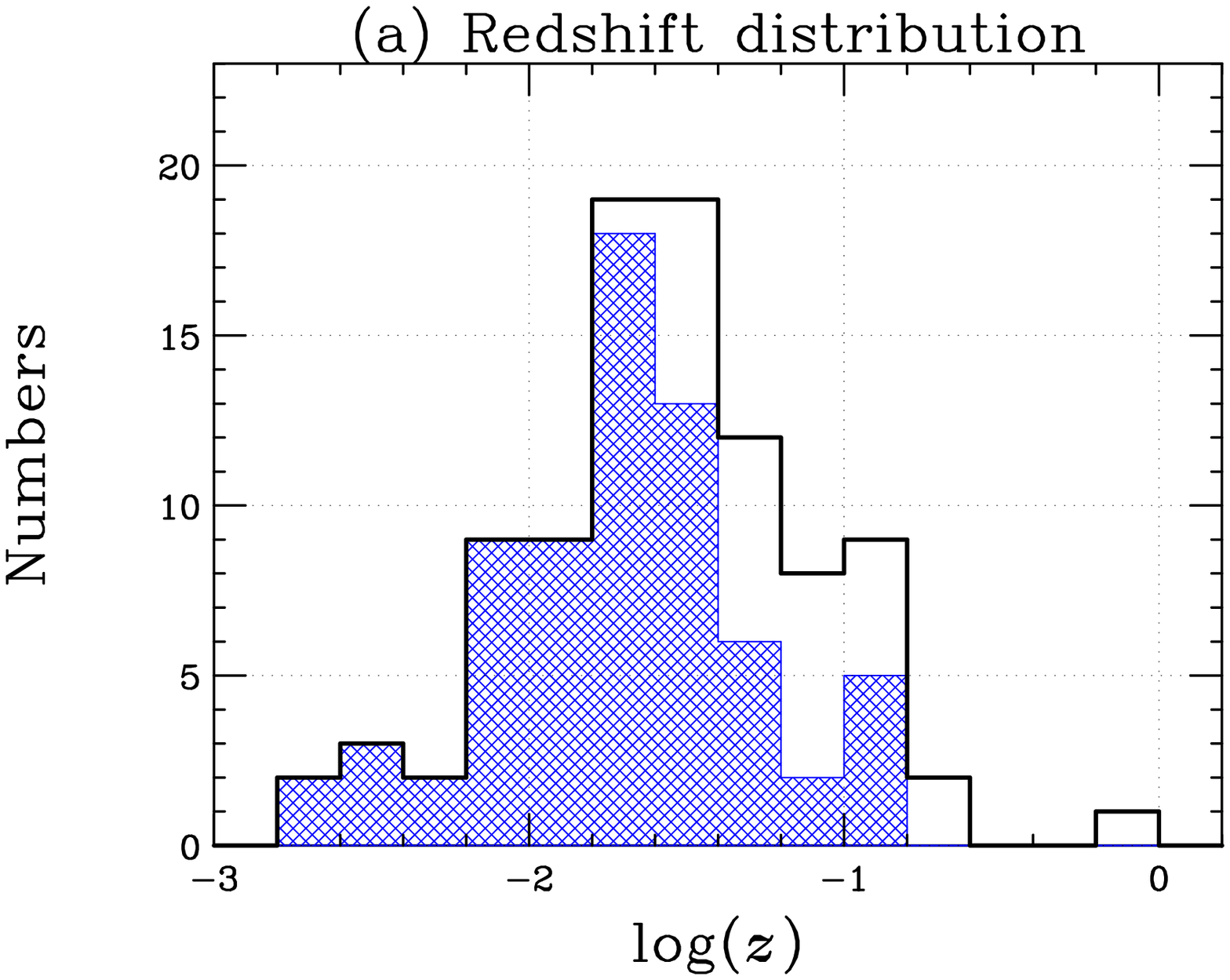}%{../fig/redshift/hist_logz.ps}
\hspace{0.5cm}
\FigureFile(75mm,75mm){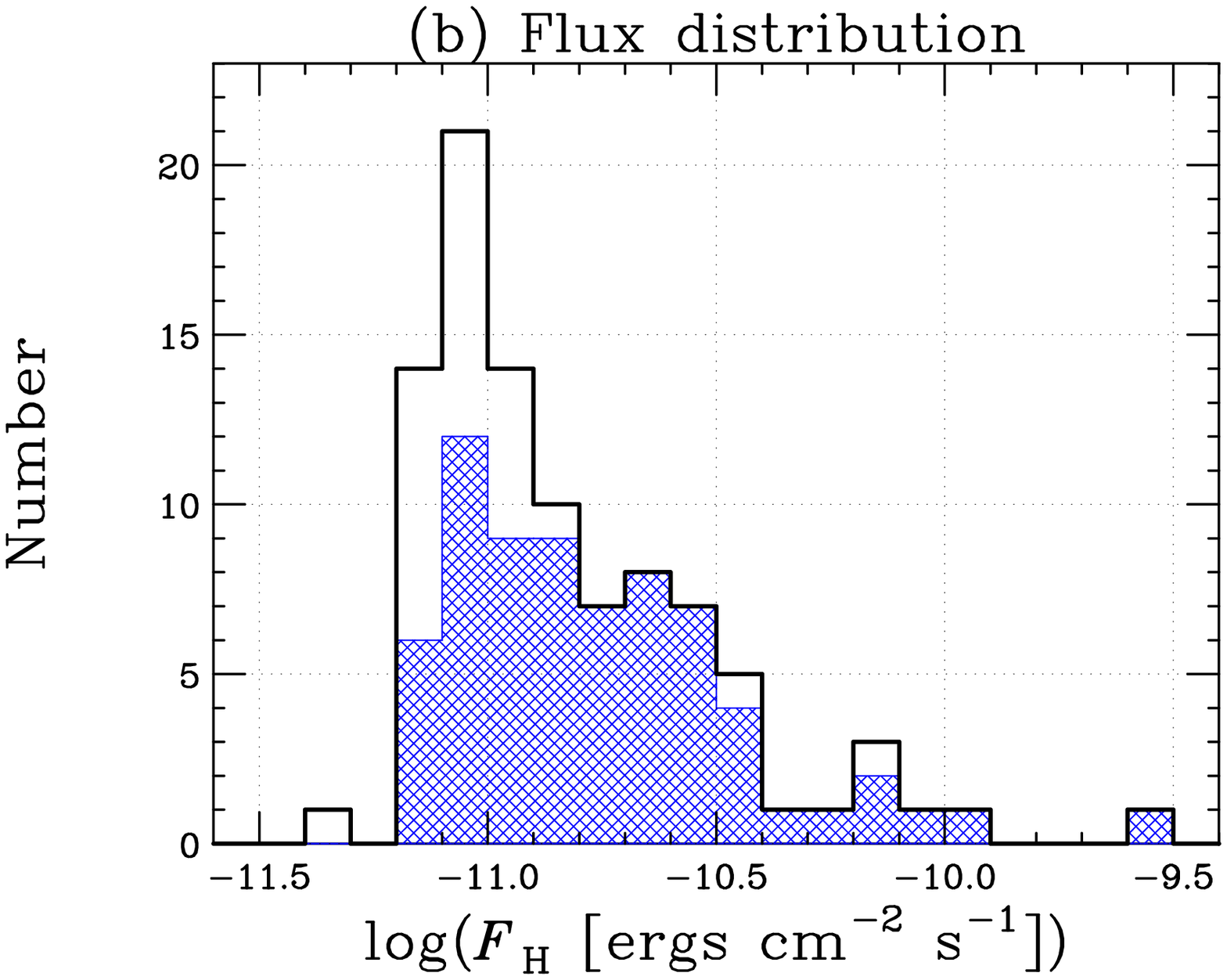}%{../fig/flux_dist/hist_logflux.ps}
\end{center}
\caption{
Distributions of the redshift $z$ (panel a) 
and $4$--$10$ keV  X-ray flux $F_{\rm H}$ (panel b).
The thick histograms indicate the distributions 
for all the Seyfert galaxies in the 2MAXI catalog,
while the hatched histograms show the distributions 
for those with an AKARI counterpart.  
}
\label{fig:distribution}
\end{figure*}

%%%%%%%%%%%%%
% Section 2 %
%%%%%%%%%%%%%
% \clearpage
\section{Sample}        %=============================================
\label{sec:sample}
\subsection{The second MAXI all-sky X-ray source catalog}    %---------------------
\label{sec:2MAXI}
The 2MAXI catalog \citep{2MAXI}\footnote{Electrically available at \\
{\tt http://vizier.cfa.harvard.edu/viz-bin/VizieR?-source=J/ApJS/207/36}.}
was constructed from all-sky X-ray survey data 
accumulated with the gas slit camera \citep{MAXI/GSC1,MAXI/GSC2} onboard MAXI 
in the first 37 months from 2009 September to 2012 October. 
The catalog lists 500 X-ray sources detected 
at high Galactic latitude of $|b| \ge 10 \degree$
with a significance of $>7\sigma$ in the $4$ -- $10$ keV range. 
Its $7\sigma$ detection sensitivity, 
$7.5 \times 10^{-12}$ erg cm$^{-2}$ s$^{-1}$ in $4$ -- $10$ keV 
for about half of the sky, 
is highest among those of previous all-sky X-ray surveys 
in the similar energy range.
The X-ray flux of the faintest source in the catalog 
is measured as $4.7 \times 10^{-12}$ erg cm$^{-2}$ s$^{-1}$. 

\citet{2MAXI} cross-matched all the 2MAXI X-ray sources
to previous X-ray source catalogs, 
including the first MAXI catalog \citep{1MAXI},
the meta-catalog of X-ray detected clusters of galaxies \citep{MCXC}, 
and the Swift/BAT 70-month catalog \citep{BAT70}.
As a result, 
they reported that $292$ 2MAXI sources have a probable single counterpart. 
In the following, 
we assume that these are real counterparts to the 2MAXI sources. 
In addition, 
multiple candidate counterparts were eventually found for $4$ sources,
mostly due to the moderate position uncertainty of the 2MAXI catalog
(90\% error radius of $\sim 0.5\degree$ for $7\sigma$ sources;
\cite{2MAXI}).
Thus, the current completeness of source identification  
to the 2MAXI catalog is about $\sim 58$\%.
For the remaining 204 sources, identification studies are ongoing 
(private communication with the MAXI team). 

In the present study, all the active galactic nuclei
tabulated in the 2MAXI catalog were picked up;
these consist of 95 Seyfert galaxies, 5 quasars, and 
15 blazars including BL Lacertae objects. 
We noticed that all of these sources are listed 
in the Swift/BAT 70-month catalog.
We refer to the 2MAXI catalog (and references therein)
for their source name, the 3-year averaged X-ray fluxes 
in the $3$--$4$ keV and $4$--$10$ keV ranges
($F_{\rm S}$ and $F_{\rm H}$, respectively), 
X-ray hardness ratio between these bands $HR$ 
(its definition is discussed in \S\ref{sec:color-color}), 
optical position, redshift $z$, and object type (optical classification). 

%---------%
% Table 2 %
%---------%
\begin{table}[t]
\caption{Statistics of Source Identification}
\label{tab:sourceID}
\begin{center}
\begin{tabular}{llll}
\hline\hline %==================================================
Type                              &  Sy1  & Sy2  & Sy1+Sy2 \\
\hline       %--------------------------------------------------
$N_{{\rm X}}$\footnotemark[$*$]   &  68   & 27   & 95  \\
$N_{IR}$\footnotemark[$\dagger$]  &  48   & 21   & 69 \\
$N_{9}$\footnotemark[$\ddagger$]  &  36   & 16   & 52 \\
$N_{18}$\footnotemark[$\ddagger$] &  38   & 19   & 57 \\
$N_{90}$\footnotemark[$\ddagger$] &  30   & 20   & 50 \\
\hline       %--------------------------------------------------
\multicolumn{4}{@{}l@{}}{\hbox to 0pt{\parbox{50mm}{\footnotesize
\vspace{0.2cm}
\par\noindent\footnotemark[$*$] 
Number of the Seyfert galaxies listed in the 2MAXI catalog.
\par\noindent\footnotemark[$\dagger$] 
Number of the 2MAXI Seyfert galaxies,
that have an AKARI counterpart 
in at least one of the three photometric bands.
\par\noindent\footnotemark[$\ddagger$] 
Number of the 2MAXI Seyfert galaxies,
detected at $9$ $\mu$m ($N_{9}$), $18~\mu$m ($N_{18}$) or $90$ $\mu$m ($N_{90}$).
}\hss}}
\end{tabular}
\end{center}
\end{table}

\subsection{The AKARI point source catalog}      %--------------------
\label{sec:AKARI/PSC} 
The AKARI/PSC is electrically available 
at the AKARI Catalogue Archive Server \citep{AKARI-CAS}
\footnote{{\tt http://darts.isas.jaxa.jp/astro/akari/cas.html}}.
The catalog contains $870,973$ mid-IR sources \citep{IRC_Catalog}
and $427,071$ far-IR ones \citep{FIS_Catalog}, 
detected with the IRC and the FIS, respectively.
The IRC is equipped with $2$ photometric bands 
at the effective wavelength of $\lambda = 9$ $\mu$m and $18$ $\mu$m.
The IRC sensitivity for an $80$\% detection completeness is  
0.12 Jy and 0.22 Jy at $9~\mu$m and $18~\mu$m, respectively. 
The FIS has $4$ photometric bands 
centered at $\lambda = 65$ $\mu$m, $90$ $\mu$m, $140$ $\mu$m, and $160$ $\mu$m.
In the present study, 
we refer only to $90$ $\mu$m FIS sources, 
since the sensitivity at this band (the $80$\% completeness limit of 0.43 Jy)
is nearly an order of magnitude better than those at the other 3 bands. 
The typical position accuracy of the IRC and FIS sources 
is $\sim 3\arcsec$ and $\sim6\arcsec$, respectively. 

A quality flag, 
indicating a reliability of the source detection and flux determination,
is assigned to each AKARI source for the individual photometric bands. 
According to the recommendation from the AKARI team,
we adopted only the AKARI sources with a quality flag of 3,
which assures a high flux accuracy. 
The same criterion was commonly adopted in similar studies 
(e.g., \cite{Swift/BAT-AKARI_AGN1,Swift/BAT-AKARI_AGN2,CT-AGN_XMM-AKARI}).

%--------------------------%
% Figure 3: logLIR - logLX %
%--------------------------%
\begin{figure*}[t]
\centerline{
\FigureFile(150mm,150mm){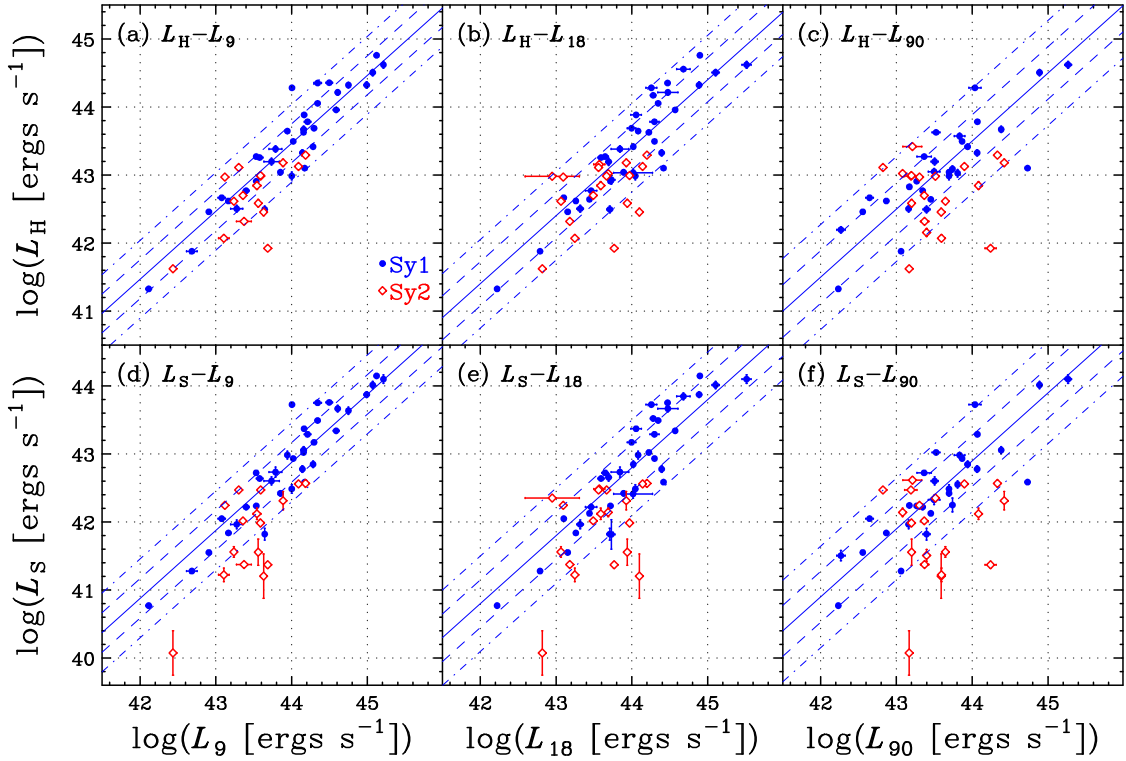}}
% {../fig/logLIR-logLX/fig/logLX-logLIR.ps}
\vspace{0.2cm}
\caption{Relation between the observed X-ray and IR luminosities;  
(a) $L_{\rm H}$ -- $L_{9}$,
(b) $L_{\rm H}$ -- $L_{18}$,
(c) $L_{\rm H}$ -- $L_{90}$,
(d) $L_{\rm S}$ -- $L_{9}$,
(e) $L_{\rm S}$ -- $L_{18}$, and 
(f) $L_{\rm S}$ -- $L_{90}$.
The Sy1 and Sy2 galaxies are plotted with the filled blue circles 
and open red diamonds, respectively.
The logarithmic average of the X-ray to IR luminosity ratio 
for the Sy1 galaxies is shown with the solid line in each panel, 
while the $1 \sigma_{\rm r}$ and $2 \sigma_{\rm r}$ ranges are indicated 
by the dashed and dash-dotted lines, respectively. 
}
\label{fig:LIR-LX}
\end{figure*}

%-----------------------%
% Figure 4: log(LX/LIR) %
%-----------------------%
\begin{figure*}[t]
\centerline{
\FigureFile(150mm,150mm){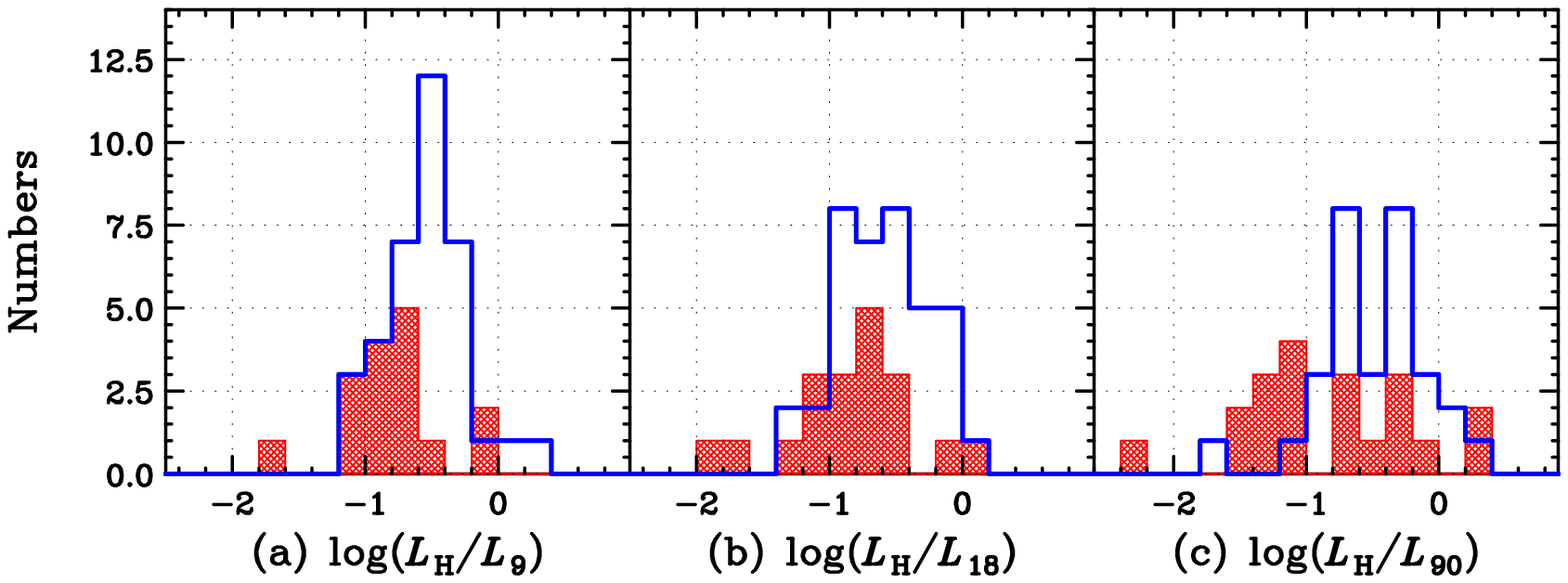}}
% {../fig/Ratio_logLIR-logLX/fig/bin2/hist_log_Ratio_LH-LIR.ps}
\vspace{0.2cm}
\centerline{
\FigureFile(150mm,150mm){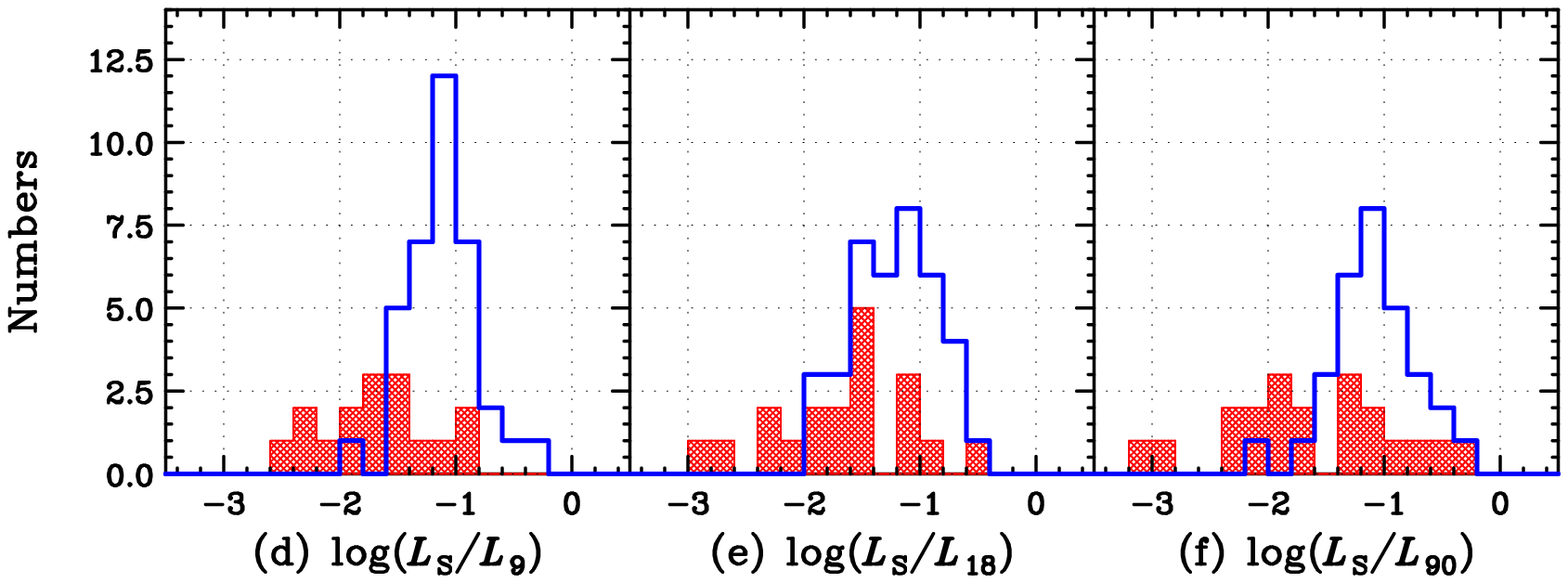}}
% {../fig/Ratio_logLIR-logLX/fig/bin2/hist_log_Ratio_LS-LIR.ps}
\vspace{1.2cm}
\caption{Distributions of the absorption-inclusive 
X-ray to IR luminosity ratio in the logarithmic space;
(a) $\log(L_{\rm H}/L_{\rm 9})$,
(b) $\log(L_{\rm H}/L_{\rm 18})$,
(c) $\log(L_{\rm H}/L_{\rm 90})$,
(d) $\log(L_{\rm S}/L_{\rm 9})$,
(e) $\log(L_{\rm S}/L_{\rm 18})$,
and 
(f) $\log(L_{\rm S}/L_{\rm 90})$.
The distributions of the Sy1 and Sy2  galaxies 
are indicated with the thick blue and hatched red 
histograms respectively. 
}
\label{fig:Ratio_LX-LIR}
\end{figure*}

\subsection{Source identification} %====================================
\label{sec:sourceID}
We searched the AKARI/PSC for IR counterparts 
to the active galaxies selected from the 2MAXI catalog.
For the source identification,
we referred to the optical position of the individual 2MAXI sources,
instead of their position determined with MAXI. 
A search radius of $10\arcsec$ and $20\arcsec$ 
was adopted for the IRC and FIS catalog, respectively, 
since these values nearly correspond to their $3\sigma$ position accuracy.
The same angular threshold was widely imposed in previous studies
(e.g., \cite{Swift/BAT-AKARI_AGN1}).

The result of the source identification is summarized 
in table \ref{tab:catalog}, 
where the X-ray properties ($F_{\rm H}$, $F_{\rm S}$, and $HR$) 
and IR flux densities 
at $9$ $\mu$m, $18$ $\mu$m, and $90$ $\mu$m 
($f_{9}$, $f_{18}$, and $f_{90}$ respectively) 
are tabulated for those listed in both the 2MAXI catalog and AKARI/PSC, 
together with their optical information.
Figure \ref{fig:dist_AngSep} plots the distribution of the angular separation 
between the 2MAXI sources and their AKARI counterparts.
As shown with the hatched histogram, 
the IRC counterparts are concentrated at a relatively narrow  range 
with a separation of $\lesssim 6 \arcmin$.
Due to the slightly worse angular resolution 
at the longer wavelength, the distribution of the FIS sources
(up to $\sim 17\arcmin$ as indicated by the thick histogram)
is wider than that of the IRC sources. 
These trends are qualitatively
consistent with the previous study by \citet{Swift/BAT-AKARI_AGN2}.

No AKARI counterpart was found to the 5 quasars in the 2MAXI catalog. 
This is reasonable 
because these quasars are a relatively high-redshift and faint source
($z > 0.17$ and $F_{\rm H} < 1.5 \times 10^{-11}$ erg cm$^{-2}$ s$^{-1}$; 
see figure \ref{fig:distribution}).
Among 15 blazars, only one BL Lacertae object PKS 0521$-$36, 
corresponding to the MAXI X-ray source 2MAXI J0523$-$363 
with a hard and soft X-ray flux of 
$F_{\rm H} = (12.7 \pm 1.2) \times 10^{-12} $ erg cm$^{-2}$ s$^{-1}$and 
$F_{\rm S} = (3.5 \pm 0.4) \times 10^{-12} $ erg cm$^{-2}$ s$^{-1}$ respectively, 
was identified with an AKARI source 
with $9$ $\mu$m and $18$ $\mu$m IR flux densities of
$f_{9}=96.6\pm0.1$ mJy and $f_{18}=216\pm20$ mJy, respectively. 
Therefore, in the statistical argument below, 
we deal only with Seyfert galaxies. 

Statistics of the source matching for the Seyfert galaxies
between the 2MAXI catalog and AKARI/PSC 
are summarized in Table \ref{tab:sourceID}.
Among the 95 2MAXI Seyfert galaxies, 
69 ones ($\sim 73$\%) were successfully identified with an AKARI source 
at least at one of the three photometric bands.
The number of the 2MAXI Seyfert galaxies 
detected at $9$ $\mu$m, $18$ $\mu$m, and $90$ $\mu$m 
is $N_{\rm 9} = 52$, $N_{\rm 18} = 57$, and $N_{\rm 90} = 50$, respectively. 
Out of the 68 type-1 Seyfert (Sy1) galaxies, 
including those optically categorized into Sy1.2 and Sy1.5 sources, 
$48$ ($\sim 71$\%) are found to have an AKARI counterpart
($N_{\rm 9} = 36$, $N_{\rm 18} = 38$, and $N_{\rm 90} = 30$),
while 21 out of 27 type-2 Seyfert (Sy2) galaxies ($\sim 78$\%),
including Sy1.8 and Sy1.9 objects, were detected with AKARI 
($N_{\rm 9} = 16$, $N_{\rm 18} = 19$, and $N_{\rm 90} = 20$).

The redshift distribution of the 2MAXI Seyfert galaxies is displayed 
in the panel (a) of figure \ref{fig:distribution}.
Their average redshift was evaluated as $\overline{z} \simeq 0.048$.
When we focus on the objects with an AKARI counterpart, 
the average redshift reduces to $\overline{z} \simeq 0.03$.
All the 2MAXI Seyfert galaxies located 
at $z < 0.023$ (41 ones) are detected with AKARI. 
In contrast, only 5 Seyfert galaxies (out of 12 ones) at $z > 0.1$
are found to have an AKARI counterpart, 
with PKS 1549$-$79 (Sy1) being the most distant one ($z = 0.1501$).
Thus, our final sample 
is limited to local sources mainly at $z \lesssim 0.1$. 

The $4$--$10$ X-ray flux, $F_{\rm H}$, of the 2MAXI Seyfert galaxies 
is distributed as shown in the panel (b) of figure \ref{fig:distribution}.
The X-ray brightest source in the sample is Centaurus A, 
with an X-ray flux of 
$F_{\rm H} = 3.00 \times 10^{-10}$ erg cm$^{-2}$ s$^{-1}$ 
averaged over the 3 years.
It is reasonable that the objects, that are not detected with AKARI, 
are mainly the fainter ones with an X-ray flux of 
$F_{\rm H} \lesssim 1\times 10^{-11}$ erg cm$^{-2}$ s$^{-1}$. 
Above this flux threshold, 
$59$ out of $67$ objects ($\sim 86$\%) were found 
to coincide with an AKARI source. 
Among the objects without any AKARI counterpart,
the X-ray brightest one is the Sy1.9 galaxy 2MASX J09235371$-$3141305 
with a hard-band X-ray flux 
of $F_{\rm H} = 6.35 \times 10^{-11}$ erg cm$^{-2}$ s$^{-1}$.

\section{Results} %=============================================================
\subsection{Relation between the X-ray and IR luminosities} %------------
\label{sec:LX-LIR}
In figure \ref{fig:LIR-LX},
the soft- and hard-band X-ray luminosities,
 $L_{\rm S}$ and $L_{\rm H}$ respectively, 
of the 69 Seyfert galaxies tabulated in both the 2MAXI and AKARI catalogs,
are plotted against their IR ones at $9$ $\mu$m, $18$ $\mu$m and $90$ $\mu$m,
$L_{\rm 9}$, $L_{\rm 18}$ and $L_{\rm 90}$.  
% \textcolor{blue}{
The solid line on each panel corresponds to 
the average of the X-ray-to-IR luminosity ratio among the Sy1 galaxies 
evaluated in the logarithmic space,
while the dashed and dash-dotted lines show
its $1\sigma_{\rm r}$ and $2\sigma_{\rm r}$ ranges, respectively,
where $\sigma_{\rm r}$ indicates the standard deviation 
of the logarithmic luminosity ratio.
%}
Here, the X-ray flux observed with MAXI was simply converted 
into the luminosity as $L_i = 4 \pi D_{\rm L}^2 F_i$ 
($i = {\rm H}$ and ${\rm S}$ for the hard and soft band respectively).
The source redshift $z$ was transformed to the luminosity distance $D_{\rm L}$,
by assuming the cosmological constants of $H_{\rm 0} = 71$ km s$^{-1}$ Mpc$^{-1}$, 
$\Omega_{\rm M}=0.27$, and $\Omega_{\rm \Lambda}=0.73$.
The monochromatic IR luminosity was calculated 
from the flux density at each photometric band 
as $L_\lambda = 4 \pi D_{\rm L}^2 \nu_\lambda f_\lambda $,
where $\nu_{\rm \lambda}$ is the representative frequency 
of the AKARI photometric bands
(i.e.,  $\lambda = 9$ $\mu$m, $18$ $\mu$m, and $90$ $\mu$m). 
We neglected the so-called $K$-correction 
because the sample is limited to the low-redshift sources,
as shown in figure \ref{fig:distribution}.
Even for the most distant source in the sample, PKS 1549$-$79 located at $z = 0.1501$,
the effect of the K-correction on its luminosity is evaluated as less than a few \%.
  
Figure \ref{fig:LIR-LX} suggests that 
the X-ray luminosities of the Sy1 galaxies in the 2MAXI-AKARI sample 
linearly correlates with the IR ones in the logarithmic space,
while the correlation appears to be rather vague for the Sy2 sources. 
In order to quantify such a correlation, 
we calculated the Spearman's rank correlation coefficient ($\rho_{\rm L}$), 
between the X-ray and IR luminosities in the logarithmic space.
This coefficient was commonly adopted for similar studies 
(e.g., \cite{Swift/BAT-AKARI_AGN1,Swift/BAT-AKARI_AGN2}).
The result is summarized in table \ref{tab:correlation_L}.
A tight correlation was confirmed for the Sy1 galaxies 
with $\rho_{\rm L}\sim 0.9$. 
In contrast,
the X-ray luminosities of the Sy2 objects are found 
to exhibit a moderate correlation to the mid-IR luminosities 
($\rho_{\rm L} \simeq 0.3$--$0.5$ to $\log(L_{\rm 9})$ and $\log(L_{\rm 18})$), 
and no meaningful one to the far-IR luminosity  
($\rho_{\rm L}\sim 0$ to $\log(L_{\rm 90})$).
After the Sy2 galaxies are added to the Sy1 ones 
(indicated as Sy1+Sy2 in table \ref{tab:correlation_L}), 
the coefficients to $\log(L_{\rm 9})$ and $\log(L_{\rm 18})$ still indicate 
a strong correlation ($\rho_{\rm L}\gtrsim 0.8$),
while those to $\log(L_{\rm 9})$ decreased to 
a moderate value ($\rho_{\rm L}\sim 0.6$). 

We also evaluated the Spearman's rank correlation coefficients
between the logarithms of the observed X-ray fluxes and IR flux densities 
($\rho_{\rm F}$), and tabulate them in table \ref{tab:correlation_F}.
In the case of flux-limited samples, 
the flux correlation is useful,
since it is free from artifacts due to the source redshift. 
A comparison between the results 
in tables \ref{tab:correlation_L} and \ref{tab:correlation_F} pointed out 
that $\rho_{\rm F}$ exhibits a similar trend to $\rho_{\rm L}$,
except for the fact that 
the tight luminosity correlation suggested for the Sy1 and Sy1+Sy2 categories
has reduced to a moderate one in the flux space ($\rho_{\rm F}=0.4$--$0.6$).
We think that this is probably because the observed X-ray flux range  
(typically $\log(F_{\rm H}) = -11.0$--$-10.5$
as shown in figure \ref{fig:distribution}) 
is only comparable to the dispersion of the X-ray-to-IR flux/luminosity ratio
(see below).

Figure \ref{fig:Ratio_LX-LIR} plots 
the distribution of the X-ray-to-IR luminosity ratio 
for the 2MAXI-AKARI sample in the logarithmic space.
The Sy1 galaxies are found to be distributed 
in a relatively narrow range with $\sigma_{\rm r} \simeq 0.3 $--$0.4$
(corresponding to a factor of $2$--$3$ in the linear space).
This range is found to be similar to those reported 
in the previous studies (e.g., \cite{Swift/BAT-AKARI_AGN1,Swift/BAT-AKARI_AGN2}). 
%\textcolor{blue}{
In contrast, 
the observed X-ray luminosity of the Sy2 galaxies 
tends to be lower than that of the Sy1 galaxies with a similar IR luminosity.
This is clearly visualized in figure \ref{fig:LIR-LX}
where the majority of the Sy2 galaxies are located below the solid line,
indicating the logarithmic average of the X-ray-to-IR luminosity ratio 
for the Sy1 galaxies.
%}
This trend is more prominent in the soft X-ray band,
as is clearly indicated in figure \ref{fig:Ratio_LX-LIR}.
As we discuss in \S \ref{sec:NH-correction},
this tendency is basically attributable to X-ray absorption.  

%-----------------------%
% Figure 5: XRH-IRcolor %
%-----------------------%
\begin{figure*}[t]
\centerline{
\FigureFile(150mm,150mm){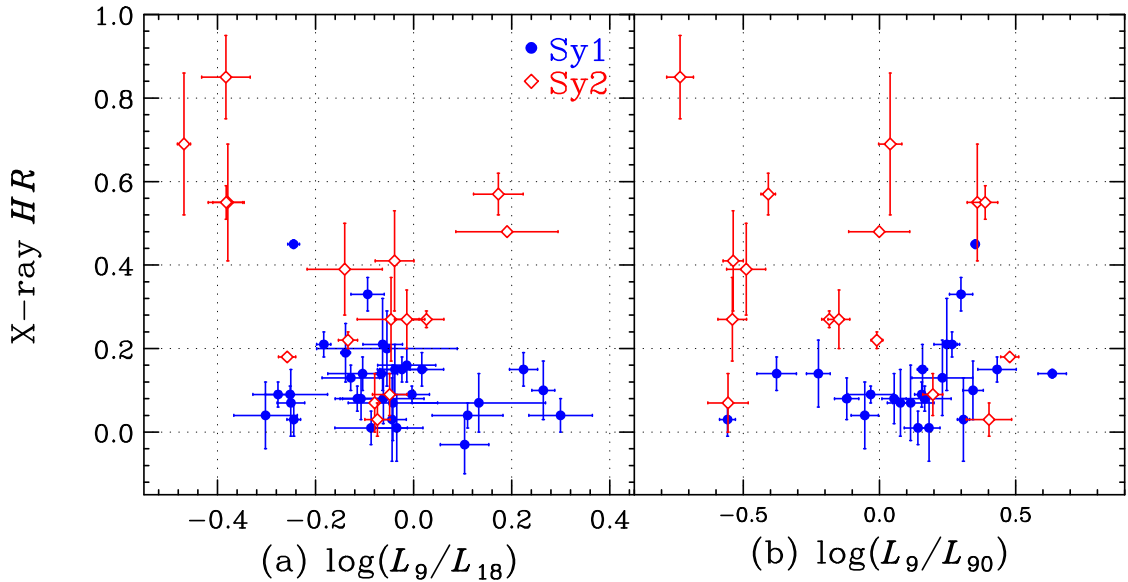}}
%{../fig/XHR-IRcolor/fig/XHR-IRcolor-3.ps}
\caption{
The relation between the X-ray hardness $HR$ and IR colors.
In panels (a) and (b),
$HR$ is plotted against $\log(L_{\rm 9}/L_{\rm 18})$ 
and $\log(L_{\rm 9}/L_{\rm 90})$, respectively.
The filled blue circles indicate the Sy1 galaxies,
while the open red diamonds point the Sy2 galaxies.}
\label{fig:color}
\end{figure*}
%---------------------------------%
% Figure 6: IR color distribution %
%---------------------------------%
\begin{figure*}[t]
\centerline{
\FigureFile(120mm,120mm){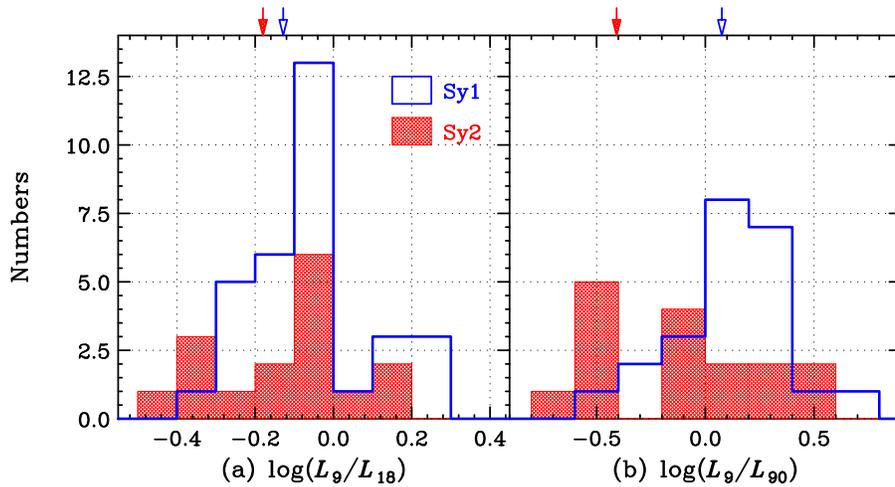}}
% {../fig/XHR-IRcolor/fig/hist_IRcolor-3.ps}
\caption{Distributions of the IR colors 
among the 2MAXI-AKARI Seyfert galaxies.
The panels (a) and (b) display 
the distributions of $\log(L_{\rm 9}/L_{\rm 18})$
and $\log(L_{\rm 9}/L_{\rm 90})$, respectively.
The thick blue and hatched red histograms indicate 
the distributions of the Sy1 and Sy2 galaxies, 
respectively.
The open blue and filled red arrows on the top of the figure 
represent the mean IR color of the Sy1 and Sy2 galaxies, respectively,
in the 9-month Swift/BAT sample, 
estimated from \citet{Swift/BAT-AKARI_AGN2}.}
\label{fig:IRcolor_hist}
\end{figure*}

\subsection{Color-color plot} %-------------------------------
\label{sec:color-color}
In figure \ref{fig:color}, 
the X-ray hardness ratio $HR$, representing the X-ray spectral color, 
is plotted against the IR colors
$L_{\rm 9}/L_{\rm 18}$ and $L_{\rm 18}/L_{\rm 90}$
for the 2MAXI-AKARI Seyfert sample.
The X-ray hardness, taken from \citet{2MAXI}, is defined as 
$HR= (F'_{\rm H} - F'_{\rm S}) / (F'_{\rm H} + F'_{\rm S}) $,
where
$F'_{\rm H} = F_{\rm H}/F_{\rm H,C}$ and $F'_{\rm S} = F_{\rm S}/F_{\rm S,C}$
represent the X-ray fluxes normalized to the Crab ones 
in the hard and soft bands, respectively 
($F_{\rm H,C} = 1.21 \times 10^{-8}$ erg cm$^{-2}$ s$^{-1}$ and  
$F_{\rm S,C} = 3.98 \times 10^{-9}$ erg cm$^{-2}$ s$^{-1}$). 
This means that a Crab-like X-ray spectrum 
with a photon index of $\Gamma = 2.1$ and $N_{\rm H} = 3.5 \times 10^{21}$ cm$^{-2}$
\citep{CrabSpec} corresponds to $HR=0$. 

Figure \ref{fig:color} 
%\textcolor{blue}{
helps us to
%}
discriminate the Sy1 galaxies from the Sy2 ones.
On the color-color plots, 
the Sy1 objects tend to concentrate in a relatively narrow area 
represented by $0 \lesssim HR \lesssim 0.2$ 
and $|\log(L_{\rm 9}/L_{\rm 18})| \lesssim 0.3$,
or  $|\log(L_{\rm 9}/L_{\rm 90})| \lesssim 0.6$.
Compared to the Sy1 sources, 
the Sy2 ones exhibit a rather hard X-ray spectrum, 
with a typical X-ray hardness of $HR \gtrsim 0.2$.

We accumulated the IR-color distributions of the sample,
as shown in figure \ref{fig:IRcolor_hist}.
The arrows on the top of each panel indicate the IR colors,
which were estimated from the average spectral energy distribution 
of the Sy1 and Sy2 galaxies in the 9-month Swift/BAT sample 
taken from \citet{Swift/BAT-AKARI_AGN2}.
Thus, the IR color of the 2MAXI sample is 
consistent to that of the 9-month Swift/BAT sample.
We found no significant difference in the mid-IR color, 
$\log(L_{\rm 9}/L_{\rm 18})$, between the two Seyfert classes, 
although there are a few relatively red Sy2 galaxies 
with Mrk 3 being the reddest one with $\log(L_{\rm 9}/L_{\rm 18}) = -0.47$.
In contrast, the distribution of the mid-to-far IR color,
$\log(L_{\rm 18}/L_{\rm 90})$,
seems to rather differ between the Sy1 and Sy2 galaxies 
in the sense that the Sy2 galaxies tend to show a redder 
mid-to-far IR color with $\log(L_{\rm 18}/L_{\rm 90}) \lesssim 0$. 
From the K-S test, 
the probability of the difference between the two source categories
 was estimated as $\sim 95$\% 

%-------------------------------------%
% Figure 7: Absorption Correction %
%-------------------------------------%
\begin{figure}[t]
\centerline{
\FigureFile(80mm,80mm){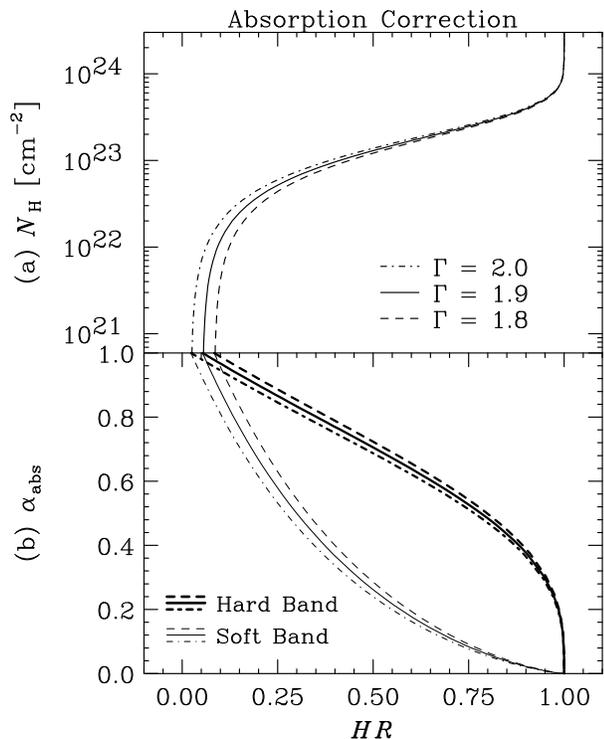}}
% {../fig/HR-NH_col/fig/HR-NH_col.ps}
\caption{
(a) The line-of-sight hydrogen column density, $N_{\rm H}$,
plotted as a function of the X-ray hardness, $HR$, derived with MAXI.
The model predictions, 
corresponding to the intrinsic photon index of $\Gamma = 1.8, 1.9$, and $2.0$,
are indicated with the dashed, solid and dash-dotted lines, respectively.
(b) The absorbed-to-intrinsic flux ratio $\alpha_{\rm abs}$
in the soft (thin lines) and hard (thick lines) bands, plotted against $HR$.
}
\label{fig:HR-NH_cor}
\end{figure}

\section{Discussion} %=========================================================
\subsection{Absorption column density estimated from $HR$} %--------------------------------
\label{sec:NH-correction}
The X-ray continuum from Seyfert galaxies in the $2$--$10$ keV range
is thought to be basically dominated by the direct nuclear X-ray emission,
which is absorbed by their dust torus, except for Compton-thick sources 
(e.g., \cite{XRB_Gilli,XRB_Ueda}).
Because X-ray photons in the soft band are more easily subjected 
to absorption than those in the hard band,
an object with higher absorption column density 
is predicted to exhibit a harder X-ray spectrum 
(at least in the Compton-thin regime).
Therefore, the higher X-ray hardness observed from the Sy2 objects,
which is clearly visualized in figure \ref{fig:color}, 
is naturally ascribed to the X-ray absorption. 

From the X-ray hardness ratio derived with MAXI,
we can roughly estimate the absorbing column density of the dust torus
\citep{LumFunc_MAXI}.
Here, we simply assume that the dominant X-ray spectral component 
from the Seyfert nuclei in the MAXI energy range 
is described with an absorbed power-law model.
Panel (a) of figure \ref{fig:HR-NH_cor} 
displays the relation of the line-of-sight hydrogen column density, 
$N_{\rm H}$, to the MAXI hardness, $HR$,
for some representative values of the photon index ($\Gamma = 1.8, 1.9$, and $2.0$).
This figure indicates that
the $HR$ value is sensitive to the X-ray absorption 
in the range of $N_{\rm H} = 10^{22}$--$5\times10^{23}$ cm$^{-2}$
(i.e., in the Compton-thin regime). 
In panel (b) of figure \ref{fig:HR-NH_cor},
the ratio of the absorbed X-ray flux to the intrinsic one, 
$\alpha_{\rm abs}$, is plotted against $HR$. 
It is possible to estimate the absorption-corrected intrinsic flux/luminosity  
of the objects in our sample, by dividing $\alpha_{\rm abs}$ 
into their flux/luminosity observed with MAXI. 

In order to validate the method for absorption correction, 
we here focus on Centaurus A, 
the brightest Seyfert (Sy2) galaxy in the sample. 
%with $F_{\rm H} =  (300.2 \pm 1.8) \times 10^{-12} $ ergs s$^{-1}$ cm$^{-2}$
%\citep{2MAXI}. 
Based on the result presented in figure \ref{fig:HR-NH_cor} (a),
the hardness ratio of this object measured with MAXI,
$HR=0.48 \pm 0.1$, is converted 
into the column density of $N_{\rm H} \simeq 1.2 \times 10^{23}$ cm$^{-2}$.
From a close examination on the 0.5 -- 300 keV wide-band X-ray spectrum 
of Centaurus A obtained with Suzaku in 2005, 
\citet{CenA_Suzaku} revealed 
that its dominant power-law component in the 2 -- 10 keV range,
with a photon index of $\Gamma = 1.8$ -- $1.85$,  
is subjected to X-ray absorption 
with a column density of $N_{\rm H} = 1.5 \times 10^{23}$ cm$^{-2}$.
A similar $N_{\rm H}$ value was recently derived 
in a coordinated observation with NuStar and XMM-Newton 
performed in 2013 \citep{CenA_NuStar}. 
The hydrogen column density inferred from the X-ray hardness
with MAXI agrees with these results.
From the panel (b) of figure \ref{fig:HR-NH_cor},  
the systematic uncertainty on $\alpha_{\rm abs}$
(and hence the intrinsic luminosity estimate)
due to the difference between the MAXI and Suzaku results on $N_{\rm H}$ 
is evaluated as at most $\sim30$\%.

%--------------------------%
% Figure 8: logLIR - logLX %
%--------------------------%
\begin{figure*}[t]
\centerline{
\FigureFile(150mm,150mm){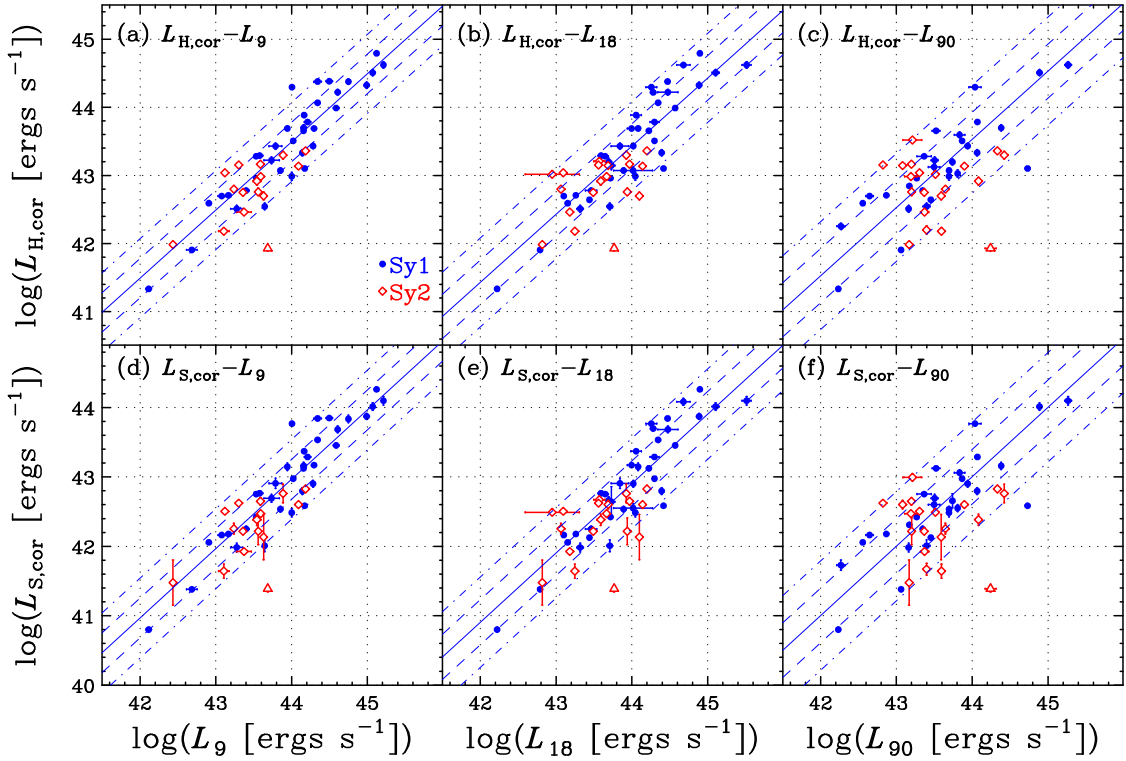}}
% {../fig/logLIR-logLX/fig/logLX-logLIR_corNH.ps}
\vspace{0.2cm}
\caption{The absorption-corrected X-ray luminosities
  plotted against the observed IR luminosities;
(a) $L_{\rm H,cor}$ -- $L_{9}$,
(b) $L_{\rm H,cor}$ -- $L_{18}$,
(c) $L_{\rm H,cor}$ -- $L_{90}$,
(d) $L_{\rm S,cor}$ -- $L_{9}$,
(e) $L_{\rm S,cor}$ -- $L_{18}$, and 
(f) $L_{\rm S,cor}$ -- $L_{90}$.
  The same symbol and color notations as for figure \ref{fig:LIR-LX}
  are adopted, except for the open triangles indicating NGC 1365. 
}
\label{fig:LIR-LX_corNH}
\end{figure*}
%--------------------------------
\subsection{X-ray-to-IR luminosity relation after the absorption correction} 
\label{sec:LIR-LX_NHcor}
By the method demonstrated in \S\ref{sec:NH-correction},
we evaluated the absorption-corrected intrinsic X-ray luminosities 
of the 2MAXI-AKARI Seyfert galaxies 
as $L_{\rm i,cor} = L_{\rm i}/\alpha_{\rm abs}$ ($i = {\rm H}$ and ${\rm S}$).  
For the absorption correction,
we adopted the typical photon index of the Seyfert galaxies,  
$\Gamma = 1.9$ (e.g., \cite{XRB_Ueda}).
No correction was performed to the sources 
with the hardness smaller than $HR=0.06$
(corresponding to $\alpha_{\rm abs} = 0.98$ and $0.99$ 
in the soft and hard bands respectively),
since we think that the absorption put only a negligible impact on these sources.
The relation between the X-ray and the IR luminosities 
after the absorption correction is displayed in figure \ref{fig:LIR-LX_corNH}.

After the absorption correction, 
we re-evaluated the X-ray-to-IR correlation coefficients, 
$\rho_{\rm L}$ and $\rho_{\rm F}$,
and summarize them in tables \ref{tab:correlation_L} and \ref{tab:correlation_F},
respectively. 
It is found that the correction put only a minor impact 
on both $\rho_{\rm L}$ and $\rho_{\rm F}$,
in comparison to those before the correction.
When we take all the Sy1 and Sy2 objects into account,
the absorption-removed X-ray luminosities are found to highly correlate 
to the $9$ $\mu$m and $18$ $\mu$m mid-IR luminosities ($\rho_{\rm L} \gtrsim 0.8$), 
while their correlation to the $90$ $\mu$m far-IR luminosity is moderate 
($\rho_{\rm L} \sim 0.6$). 
The flux correlation coefficients of the X-ray to mid-IR wavelengths 
indicate a moderate correlation ($\rho_{\rm F} = 0.4$--$0.5$), 
while those to the far-IR band 
correspond to a weak correlation ($\rho_{\rm F} \sim 0.3$). 

\begin{figure*}[t]
\centerline{
\FigureFile(150mm,150mm){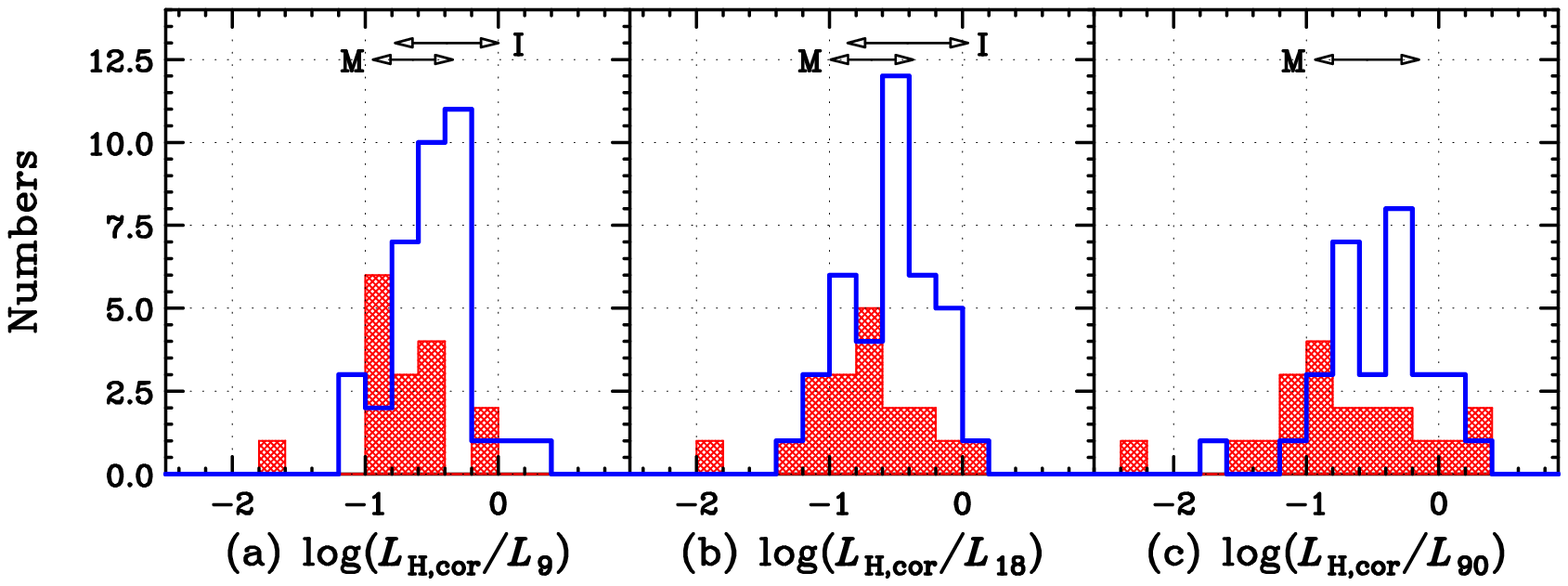}}
%{../fig/Ratio_logLIR-logLX/fig/bin2/hist_log_Ratio_LH-LIR_corNH.ps}
\vspace{0.2cm}
\centerline{
\FigureFile(150mm,150mm){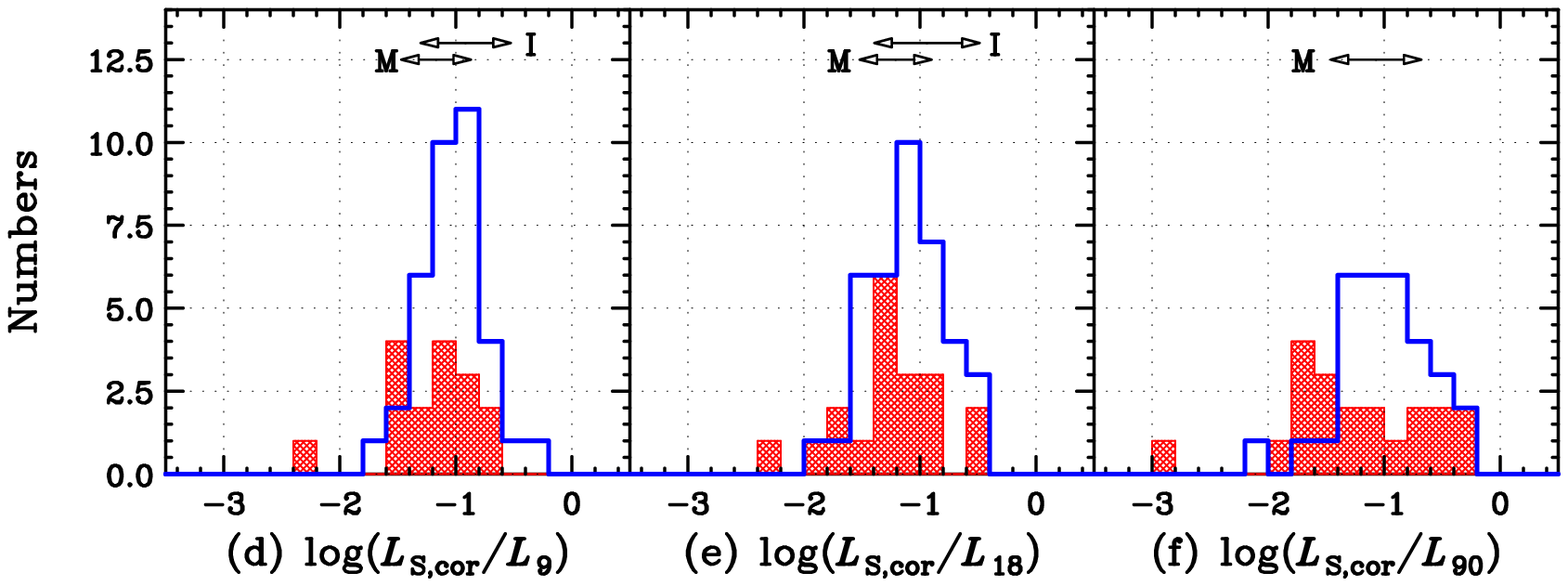}}
%{../fig/Ratio_logLIR-logLX/fig/bin2/hist_log_Ratio_LS-LIR_corNH.ps}}
\vspace{1.2cm}
\caption{Distributions of the logarithmic ratio between the X-ray and IR luminosities,
  after the X-ray absorption correction; 
(a) $\log(L_{\rm H,cor}/L_{\rm 9})$,
(b) $\log(L_{\rm H,cor}/L_{\rm 18})$,
(c) $\log(L_{\rm H,cor}/L_{\rm 90})$,
(d) $\log(L_{\rm S,cor}/L_{\rm 9})$,
(e) $\log(L_{\rm S,cor}/L_{\rm 18})$,
and 
(f) $\log(L_{\rm S,cor}/L_{\rm 90})$
The distributions among the Sy1 and Sy2 galaxies 
are indicated with the thick blue and hatched red 
histograms respectively. 
The $1\sigma_{\rm r}$ range of the luminosity ratio for the Sy1 galaxies
estimated from \citet{Swift/BAT-AKARI_AGN1} and \citet{Swift/BAT-AKARI_AGN2}
are shown with the horizontal arrows indicated as "M" and "I", respectively.
}
\label{fig:Ratio_LX-LIR_corNH}
\end{figure*}

Figure \ref{fig:Ratio_LX-LIR_corNH} shows 
the histograms of the X-ray-to-IR luminosity ratio 
in the logarithmic space, after the X-ray absorption was corrected. 
Thanks to the correction, 
the Sy2 galaxies have typically moved 
into the $2\sigma_{\rm r}$ range of the Sy1 galaxies 
(i.e., the regions 
between the dash-dotted lines on figure \ref{fig:LIR-LX_corNH}).
This result is thought to reinforce 
the X-ray-to-IR luminosity/flux correlation throughout the Sy1 and Sy2 galaxies.
%\textcolor{blue}{
For comparison, we estimated the X-ray-to-IR color 
of the 9-month and 22-month Swift/BAT samples derived in the previous studies 
by \citet{Swift/BAT-AKARI_AGN2} and \citet{Swift/BAT-AKARI_AGN1}, 
respectively.
The ratios of the $14$--$195$ keV X-ray luminosity to the IR ones 
and their errors, 
both of which are presented 
in \citet{Swift/BAT-AKARI_AGN2} and \citet{Swift/BAT-AKARI_AGN1},
were converted to match the MAXI energy range, 
by assuming a power-law like X-ray spectrum 
with a photon index of $\Gamma = 1.9$.
The horizontal arrows in figure \ref{fig:Ratio_LX-LIR_corNH} indicate
the $1\sigma_{\rm r}$ range of the X-ray-to-IR luminosity ratio, 
thus obtained, for the 9-month and 22-month Swift/BAT samples. %}
Our result from the 2MAXI sample is found to be 
consistent to the Swift/BAT results.

A number of recent researches indicate that 
the intrinsic X-ray luminosity of Seyfert galaxies tightly correlates 
with their observed IR luminosities, 
irrespective of their optical classification 
(e.g., \cite{LIR-LX_AGN,Swift/BAT-AKARI_AGN1,Swift/BAT-AKARI_AGN2}).
The correlation is widely interpreted 
under the framework of the so-called clumpy torus geometry 
(e.g., \cite{clumpy_torus}),
since the simple torus model with a smooth dust distribution 
infers a significant dust extinction to the IR luminosity 
for obscured (i.e., type-2) sources.

Figures \ref{fig:LIR-LX_corNH} and \ref{fig:Ratio_LX-LIR_corNH},
together with the correlation coefficients summarized in tables 
\ref{tab:correlation_L} and \ref{tab:correlation_F},
support the X-ray-to-IR correlation, especially that to the mid-IR band.
Basically, these mean that 
our result strengthens the clumpy torus scenario. 
Most of the previous results were usually based on 
the hard X-ray survey above 10 keV 
\citep{Swift/BAT-AKARI_AGN1,Swift/BAT-AKARI_AGN2},
or a small sample below 10 keV 
derived in a restricted sky field \citep{LIR-LX_AGN}. 
Our study successfully complements the previous ones, 
because the sample originates in the relatively unbiased all-sky X-ray survey 
below 10 keV, conducted by MAXI with the highest sensitivity.

% scatter of log(LX/L90) %
Figures \ref{fig:LIR-LX_corNH} and \ref{fig:Ratio_LX-LIR_corNH}
suggest that the dispersion of the ratio 
between the X-ray luminosity to the far-IR one $L_{\rm 90}$ 
($\sigma_{\rm r} \simeq 0.4$ for the Sy1 galaxies) 
is slightly larger than those 
to the mid-IR ones, $L_{\rm 9}$ and $L_{\rm 18}$ ($\sigma_{\rm r} \simeq 0.3$). 
In relation, 
the X-ray correlation coefficients to $\log(L_{\rm 90})$
tends to be smaller than those to  $\log(L_{\rm 9})$ and $\log(L_{\rm 18})$.
It is pointed out that within the AKARI angular resolution
non-negligible contribution to the far-IR emission 
from warm dust related to star formation activity in the host galaxy 
resulted in this larger scatter 
and lower correlation coefficients at $90~\mu$m \citep{Swift/BAT-AKARI_AGN1}. 

%---------------------------------------------------% 
% Table 3: Luminosity Correlation coefficient %
%---------------------------------------------------%
\begin{table}[t]
\caption{Summary of the Spearman's rank correlation coefficients, $\rho_{\rm L}$, 
          between the X-ray and IR luminosities.}
\label{tab:correlation_L}
\begin{center}
{\footnotesize 
\begin{tabular}{lllll}
\hline \hline     %--------------------------------------------------
     &                  &\multicolumn{3}{c}{IR luminosity} \\  \cline{3-5}
Type & X-ray luminosity & $\log(L_{9})$ & $\log(L_{18})$ & $\log(L_{90})$ \\
\hline     %--------------------------------------------------
Sy1     & $\log(L_{\rm H})$     & $0.92$ & $0.89$ & $0.88$ \\
        & $\log(L_{\rm H,cor})$ & $0.91$ & $0.89$ & $0.86$ \\
        \cline{2-5} %----------------------------------------
        & $\log(L_{\rm S})$     & $0.92$ & $0.89$ & $0.89$ \\
        & $\log(L_{\rm S,cor})$ & $0.92$ & $0.88$ & $0.87$ \\
\hline %-------------------------------------------------- 
Sy2     & $\log(L_{\rm H})$     & $0.53$ & $0.44$ & $-0.01$  \\
        & $\log(L_{\rm H,cor})$ & $0.44$ & $0.39$ & $-0.06$ \\
        \cline{2-5} %----------------------------------------
        & $\log(L_{\rm S})$     & $0.45$ & $0.27$ & $-0.01$ \\
        & $\log(L_{\rm S,cor})$ & $0.44$ & $0.39$ & $-0.06$ \\
\hline %--------------------------------------------------
Sy1+Sy2 & $\log(L_{\rm H})$     & $0.89$ & $0.82$ & $0.60$ \\
        & $\log(L_{\rm H,cor})$ & $0.89$ & $0.82$ & $0.58$ \\
        \cline{2-5} %----------------------------------------
        & $\log(L_{\rm S})$     & $0.88$ & $0.80$ & $0.59$ \\
        & $\log(L_{\rm S,cor})$ & $0.88$ & $0.82$ & $0.58$ \\
\hline %--------------------------------------------------
\end{tabular}}
\end{center}
\end{table}

%--------------------------------------------% 
% Table 4: Flux Correlation coefficient %
%--------------------------------------------%
\begin{table}[t]
\caption{Summary of the Spearman's rank correlation coefficients, $\rho_{\rm F}$, 
          between the X-ray flux and IR flux density.}
\begin{center}
\label{tab:correlation_F}
{\small 
\begin{tabular}{lllll}
\hline  \hline     %--------------------------------------------------
     &            &\multicolumn{3}{c}{IR flux density} \\ \cline{3-5} 
Type & X-ray flux &$\log(f_{9})$ & $\log(f_{18})$ & $\log(f_{90})$ \\
\hline     %--------------------------------------------------
Sy1     & $\log(F_{\rm H})$     & $0.42$ & $0.39$ & $0.40$  \\
        & $\log(F_{\rm H,cor})$ & $0.42$ & $0.39$ & $0.40$  \\
        \cline{2-5} %----------------------------------------
        & $\log(F_{\rm S})$     & $0.47$ & $0.43$ & $0.42$  \\
        & $\log(F_{\rm S,cor})$ & $0.41$ & $0.37$ & $0.39$  \\
\hline%--------------------------------------------------
Sy2     & $\log(F_{\rm H})$     & $0.53$ & $0.19$ & $-0.05$ \\
        & $\log(F_{\rm H,cor})$ & $0.46$ & $0.22$ & $-0.01$ \\
        \cline{2-5} %----------------------------------------
        & $\log(F_{\rm S})$     & $0.57$ & $0.11$ & $-0.09$ \\
        & $\log(F_{\rm S,cor})$ & $0.44$ & $0.23$ & $-0.01$ \\
\hline%--------------------------------------------------
Sy1+Sy2 & $\log(F_{\rm H})$     & $0.42$ & $0.33$ & $0.19$ \\
        & $\log(F_{\rm H,cor})$ & $0.47$ & $0.39$ & $0.25$ \\
        \cline{2-5} %----------------------------------------
        & $\log(F_{\rm S})$     & $0.34$ & $0.23$ & $0.09$ \\
        & $\log(F_{\rm S,cor})$ & $0.46$ & $0.38$ & $0.23$ \\
\hline%--------------------------------------------------
\end{tabular}}
\end{center}
\end{table}

%----------------------------------------------%
% Figure 9: logFIR - logFX after NH correction %
%----------------------------------------------%
%--------------------------------
\subsection{Behavior of Compton-thick objects} 
\label{sec:NGC1365}
Interestingly, 
we found an outlier on figures \ref{fig:LIR-LX_corNH} and \ref{fig:Ratio_LX-LIR_corNH}, which is located significantly 
outside the $2\sigma_{\rm r}$ region of the Sy1 galaxies.
This object, indicated with the red open triangles
in figure \ref{fig:LIR-LX_corNH}, is NGC 1365.
Although this source is optically classified as Sy1.8, 
it is known to exhibit occasionally a Compton-thick X-ray spectrum 
with $N_{\rm H} > 10^{24}$ cm$^{-2}$
(e.g., \cite{NGC1365_1,NGC1365_2,NGC1365_3,NGC1365_4}). 

%\textcolor{blue}{
A standard population synthesis model by \citet{XRB_Ueda} 
which was constructed from the X-ray luminosity function of active galactic nuclei
to reproduce the X-ray background spectrum
predicts the fraction of the Compton-thick objects 
as $<1$\% and $\sim6$\% in the $2$--$10$ keV and $10$--$40$ keV bands,
respectively, at the flux limit of our sample. 
The observed Compton-thick fraction (i.e., 1 out of 100) 
is in a reasonable agreement with these predictions, 
given the fact that 
our sample is essentially 
selected by MAXI and the Swift/BAT (see \S\ref{sec:2MAXI})
with an energy coverage of 4--10 keV and 14--195 keV, respectively, 
which well overlaps with the energy bands considered in \citet{XRB_Ueda}.
%}

%----------------------%
% Figure 10 : HR - log(LX/LIR) %
%----------------------%
\begin{figure*}[t]
\centerline{
\FigureFile(150mm,150mm){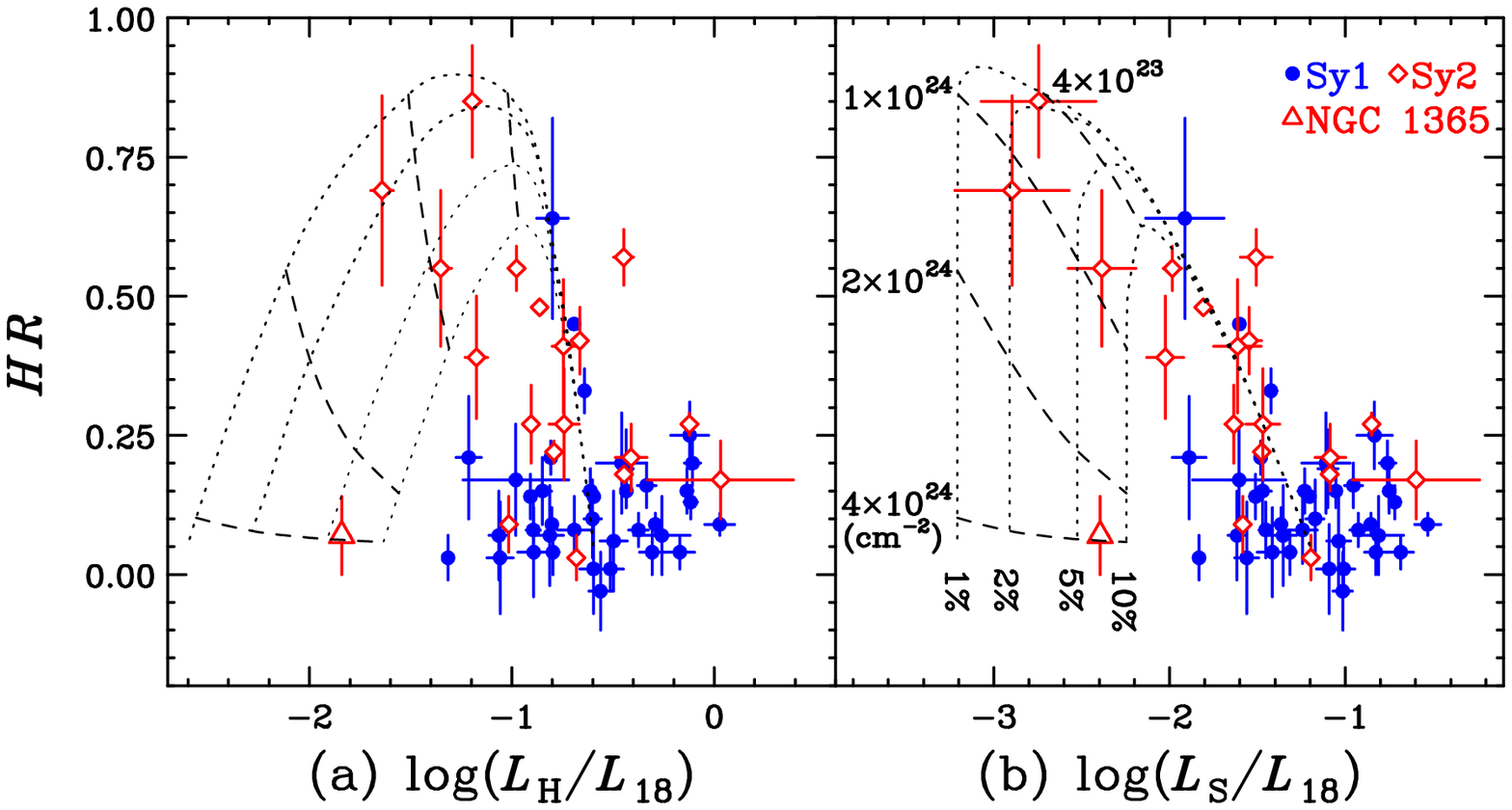}}
%{../fig/HR_logLX-LIR/fig/HR-logLX-L18.ps}}
\caption{X-ray hardness ratio $HR$ is  
plotted against the logarithm of the absorption-inclusive X-ray to IR luminosity ratios,
$\log(L_{\rm H}/L_{\rm 18})$ and $\log(L_{\rm S}/L_{\rm 18})$, 
in panels (a) and (b) respectively.
The blue filled circles and red open diamonds shows the Sy1 and Sy2 galaxies, 
respectively, while the red open triangle indicates NGC 1365.
The model prediction estimated for some representative values of 
the fraction of the scattered component 
($1$\%, $2$\%, $5$\% and $10$\%) are drawn with the dotted lines,
where the $\log(L_{\rm H}/L_{\rm 18})$ and $\log(L_{\rm S}/L_{\rm 18})$ values 
for a source with no absorption ($N_{\rm H} = 0$)
are simply assumed to be their average over the Sy1 galaxies in the sample.
The dashed lines indicate the absorption column density 
of $N_{\rm H}= 4\times 10^{23}$, $1\times 10^{24}$, 
$2\times10^{24}$, and $4\times 10^{24}$ cm$^{-2}$.}
\label{fig:HR_log(LX-LIR)}
\end{figure*}

X-ray spectrum below 10 keV of Compton-thick Seyfert galaxies 
is not dominated by the direct/absorbed emission from their nucleus. 
Instead, their dominant spectral component in this X-ray band
is though to be reflected and/or scattered ones
(e.g., \cite{XRB_Gilli,XRB_Ueda}). 
As a result, Compton-thick objects are possible 
to exhibit a relatively soft X-ray spectrum, 
in comparison to Compton-thin Sy2 galaxies
with a moderate column density (e.g., $N_{\rm H} = 10^{22}$ -- $10^{23}$ cm$^{-2}$). 
The X-ray luminosity below 10 keV of the reflected/scattered component 
is estimated to be by typically an order of magnitude 
less than the unabsorbed luminosity of the direct component. 
Therefore, for the Compton-thick sources,
the absorption correction based on the $HR$--$N_{\rm H}$ relation 
in figure \ref{fig:HR-NH_cor}
should yield a column density lower than the real value,
and thus, significantly underestimates their intrinsic luminosity. 
Using the hard X-ray catalog with Swift/BAT of nearby Seyfert galaxies,
\citet{Swift/BAT-AKARI_AGN1} reported a sign of a deficit 
in the hard X-ray luminosity from Compton-thick sources. 

The hardness ratio, $HR = 0.07 \pm 0.07$, of NGC 1365 
is within the range of those of the Sy1 galaxies, 
and thus indicates a soft X-ray spectrum in the MAXI energy range.
This $HR$ value corresponds to a low column density 
of $N_{\rm H} \sim 4 \times 10^{21}$ cm$^{-2}$ (in the case of $\Gamma = 1.9$),
in spite of its Sy2 nature. 
The X-ray-to-IR luminosity ratio of NGC 1365 
after the absorption correction is by a factor of $\gtrsim 10$ lower 
than the average value of the Sy1 galaxies, 
as shown in figure \ref{fig:LIR-LX_corNH}.
These two observational facts seem suggestive of 
the Compton-thick picture for NGC 1365.

From recent studies \citep{CT-AGN_XMM-IRAS,CT-AGN_XMM-AKARI},
the relation between the observed X-ray hardness and X-ray-to-IR color 
is proposed as an effective tool 
to pick up Compton-thick active galactic nuclei.
In panels (a) and (b) of figure \ref{fig:HR_log(LX-LIR)}, 
we plot $HR$ against the absorption-inclusive values 
of $\log(L_{\rm H}/L_{\rm 18})$ and $\log(L_{\rm S}/L_{\rm 18})$,
respectively, for the 2MAXI-AKARI sample.
On these diagrams, 
the model prediction is plotted as a function of $N_{\rm H}$ 
(from $0$ to $1\times10^{25}$ cm$^{-2}$) with the dotted lines.
The dashed lines indicate the absorption column of 
$N_{\rm H} = 4\times10^{23}$, $1\times10^{24}$, 
$2\times10^{24}$, and $4\times10^{24}$ cm$^{-2}$.
By referring to \citet{CT-AGN_XMM-AKARI},
we introduced to the absorbed power-law model,
an additional unabsorbed power-law one 
which represents the reflected/scattered component. 
The photon index of the reflected/scattered component
was presumed to be identical to the direct one 
(i.e., $\Gamma = 1.9$; see \S \ref{sec:NH-correction}).
We evaluated the model tracks,
by assuming the fraction of the scattered/reflected component 
as $1$\%, $2$\%, $5$\%, and $10$\% to 
the intrinsic luminosity of the direct component.  
The $\log(L_{\rm H}/L_{\rm 18})$ and $\log(L_{\rm S}/L_{\rm 18})$ values
of a source unaffected by absorption (i.e., $N_{\rm H} = 0$) 
were assumed to be their mean among Sy1 galaxies 
(i.e., represented by the solid lines in figure \ref{fig:LIR-LX}). 

In the Compton-thin regime ($N_{\rm H} < 10^{24}$ cm$^{-2}$),
a source with a higher $N_{\rm H}$ value is expected 
to exhibit a harder X-ray spectrum and 
a lower ratio of the soft X-ray to IR luminosities. 
In contrast, the hard-X-ray to IR color 
is relatively insensitive to the absorption.
For heavily obscured objects,
an $N_{\rm H}$ increase is predicted to result 
in a significant X-ray spectral softening
and a decrease in $\log(L_{\rm H}/L_{\rm 18})$.
The $\log(L_{\rm S}/L_{\rm 18})$ value is inferred to remain unchanged
because the direct X-ray emission in the soft band has been fully blocked 
by the Compton-thick dust torus, 
and hence, is totally dominated by the scattered/reflected component.
We clearly recognize in figure \ref{fig:HR_log(LX-LIR)}
that the behavior of the 2MAXI-AKARI Seyfert galaxies, 
qualitatively follows these trends.
Especially, we can explain the colors of NGC 1365 
by tuning the fraction of the scattered/reflected component and
the X-ray-to-IR luminosity ratio for  $N_{\rm H} = 0$,
both of which are thought to reflect the geometry of the dust torus 
around the accretion disk. 
We have, thus, re-confirmed that 
a combination of the X-ray spectral hardness below 10 keV 
and the X-ray-to-IR color is useful to distinguish 
the Compton-thick sources from the Compton-thin ones, 
without any detailed spectral analysis, or hard X-ray data above 10 keV.

Finally, it is important to mention that 
figure \ref{fig:HR_log(LX-LIR)} implies a few additional Compton-thick candidates,
which are located in the transition area 
between the Compton-thin and -thick objects, represented by 
$\log(L_{\rm H}/L_{\rm 18})\lesssim-1$ or $\log(L_{\rm S}/L_{\rm 18})\lesssim-2$
(corresponding to $N_{\rm H} > 4 \times 10^{23}$ cm$^{-2}$).
Detailed X-ray spectral analyses are considered to be 
required to find out the nature of these objects 
(e.g., \cite{CT-AGN_XMM-AKARI}), 
although we regard them as out of the scope of the present paper.

%-----------------------------------------------%
\subsection{Implication from the IR color}
\label{sec:IRcolor}

The mid-IR spectrum of Seyfert galaxies is utilized as a good probe 
into physical or geometrical properties of their dust torus,
because it is known to be dominated by emission from the dust torus 
especially in the $10$ -- $20$ $\mu$m range \citep{AGN_dust}.
It is pointed out that 
the $5$ -- $20$ $\mu$m IR spectra of nearby Sy2 galaxies are 
possibly redder than those of the Sy1 objects 
(e.g., \cite{Torus_CF}).
Based on the numerical simulation on the clumpy torus model 
(e.g., \cite{clumpy1,clumpy2}), 
these spectral characteristics are proposed to be explained 
by a possible idea that the covering factor of the dust torus 
is larger in the Sy2 sources than in the Sy1 ones \citep{Torus_CF}.
However, our result shown 
in the left panel of figure \ref{fig:IRcolor_hist}
does not seem to give a support to such a scenario,
since no clear discrepancies are found
between the Sy1 and Sy2 sources.
% although the sample is still limited. 

As briefly mentioned in \S\ref{sec:LIR-LX_NHcor},
the far-IR luminosity from the Seyfert galaxies 
measured with AKARI is thought to be 
rather contaminated by emission from warm dust 
produced by star-forming activity in their host galaxy.
Then, the mid-to-far IR luminosity ratio is regarded as 
an indicator of a relative dominance 
between the activities of the nucleus and 
star formation in the galaxies.
Using the AKARI data for the Swift/BAT sample,  
\citet{Swift/BAT-AKARI_AGN1} hinted 
a higher relative star formation activity in the Sy2 galaxies,
based on their redder $L_{\rm 9}/L_{\rm 90}$ color,
in comparison to the Sy1 objects.
We reconfirmed this trend (with a probability of $\sim 95$\% by the K-S test),
by utilizing the 2MAXI-AKARI sample, 
as shown in the right panel of figure \ref{fig:IRcolor_hist}.
In addition, no meaningful X-ray correlation to $\log(L_{90})$ 
indicated to the Sy2 galaxies  
(see tables \ref{tab:correlation_L} and \ref{tab:correlation_F})
is possible to be also related to a significant contamination 
from a higher star-formation activity in such objects.
These results suggest that the optical Seyfert classification 
is not only determined by the orientation of the dust torus 
to our line of sight,
but also by the warm-dust distribution 
connected to the star-forming activity in the host galaxy.

% \acknowledgments 
\begin{ack}
%\textcolor{blue}{
We thank the anonymous referee for her/his supportive comments 
to modify the present paper.
%}
This research is financially supported by the Ministry of Education, 
Culture, Sports, Science and Technology (MEXT) of Japan, 
through the Grant-in-Aid 24103002 (N.I.), 26247030 (T.N.) and 26400228 (Y.U.).
T.K. is supported by the Grant-in-Aid for JSPS Fellows.
We made use of the MAXI catalog (\cite{2MAXI}),
which was constructed from the MAXI data 
provided by RIKEN, JAXA and the MAXI team.
This work is based on the data taken with AKARI, 
a JAXA project, with the participation of ESA.
In advance of publication,
Dr. Ichikawa kindly provided us with the IR data of the Swift/BAT sample.
\end{ack}

%%%%%%%%%%%%%%
% References %
%%%%%%%%%%%%%%


\begin{thebibliography}{}
\bibitem[Antonucci(1993)]{AGN_Unify_1}
  Antonucci, R., 1993, \araa, 31, 473
\bibitem[Baumgartner et al.(2013)]{BAT70}
  Baumgartner, W. H., et al.
  % Tueller, J., Markwardt, C. B.,  Skinner, G. K., 
  % Barthelmy, S., Mushotzky, R. F., Evans, P. A., \& Gehrels, N.
  2013, \apjs, 207, 19
\bibitem[Burlon et al.(2011)]{Swift/BAT_CTAGN}
  Burlon, D., et al.
  % Ajello, M., Greiner, J., Comastri, A., 
  % Merloni, A., Gehrels, N.
  2011, \apj, 728, 58
\bibitem[Elitzur(2008)]{AGN_IRbump}
  Elitzur, M., 2008, New Astron., 52, 274
\bibitem[Elvis et al.(1994)]{QSO_atras}
  Elvis, M., et al. 1994, \apjs, 95, 1
  % Wilkes, B. J., McDowell, J. C., Green, R. F.,
  % Bechtold, J., Willner, S. P., Oey, M. S., Polomski, E.; Cutri, R.
\bibitem[F\"{u}rst et al.(2016)]{CenA_NuStar}
  F\"{u}rst, F., et al. 2016, \apj, accepted (arXiv:1511.01915)
\bibitem[Gandhi et al.(2009)]{LIR-LX_AGN}
 Gandhi, P., et al. 2009, \aap,  502, 457
\bibitem[Gilli et al.(2007)]{XRB_Gilli}
 Gilli, R., Comastri, A. \& Hasinger, G.,
 2007, \aap, 463, 79G
\bibitem[Hiroi et al.(2011)]{1MAXI}
  Hiroi, K., et al. 2011, \pasj, 63, S677
\bibitem[Hiroi et al.(2013)]{2MAXI}
  Hiroi, K., et al. 2013, \apjs, 207, 36
\bibitem[Ichikawa et al.(2012)]{Swift/BAT-AKARI_AGN2} 
  Ichikawa, K., et al. 
  % Ueda, Y., Terashima, Y., Oyabu, S.,
  % Gandhi, P., Matsuta, K., \& Nakagawa, T.,
  2012, \apj, 754, 45
\bibitem[Ishihara et al.(2010)]{IRC_Catalog}
  Ishihara, D., et al. 2010, A\&A, 514, A1
\bibitem[Isobe et al.(2010)]{Mrk421_MAXI_1}
  Isobe, N., et al. 2010, \pasj, 62, L55
\bibitem[Isobe et al.(2015)]{Mrk421_MAXI_2}
  Isobe, N., et al.
  %Sato, R., Ueda, Y., Hayashida, M., Shidatsu, M., Kawamuro, T.,
  %Ueno, S., Sugizaki, M., Sugimoto, J., Mihara, T.,
  %Matsuoka, M., Negoro, H.
  2015, \apj, 798, 27
\bibitem[Kawada et al.(2007)]{FIS}
  Kawada, M., et al. 2007, \pasj, 59, S389
\bibitem[Kirsch et al.(2005)]{CrabSpec}
  Kirsch, M. G., et al. 2005, SPIE, 5898, 22
\bibitem[Krolik \& Begelman(1988)]{clumpy_torus}
  Krolik, J. H., \& Begelman, M. C., 1988, \apj, 329,702 
\bibitem[Markowitz et al.(2007)]{CenA_Suzaku}
  Markowitz, A., et al. 2007, \apj, 665, 209
\bibitem[Mateos et al.(2016)]{Torus_CF}
  Mateos, S., et al. 2016, \apj, accepted (arXiv:1601.04439)
\bibitem[Matsuoka et al.(2009)]{MAXI}
  Matsuoka, M., et al. 2009, \pasj, 61, 999
\bibitem[Matsuta et al.(2012)]{Swift/BAT-AKARI_AGN1} 
  Matsuta, K., et al. 2012, \apj,  753, 104
\bibitem[Mihara et al.(2011)]{MAXI/GSC1}
  Mihara, T., et al., 2011, \pasj, 63, S623
\bibitem[Mitsuda et al.(2007)]{Suzaku}
  Mitsuda, K., et al. 2007, \pasj,  5{Suzaku}9, S1 
\bibitem[Mor et al.(2009)]{AGN_dust}
  Mor, R., Netzer, H., \& Elitzur, M.,
  2009, \apj, 705, 298
\bibitem[Murakami et al.(2007)]{AKARI}
  Murakami, H., et al. 2007, \pasj, 59, S369
\bibitem[Nenkova et al.(2008a)]{clumpy1}
  Nenkova, M., 
  % Sirocky, Matthew M., Ivezi\'{c} \'{Z}, \& Elitzur, M.
  et al. 2008b, \apj, 685, 147
\bibitem[Nenkova et al.(2008b)]{clumpy2}
  Nenkova, M.,
  % Sirocky, Matthew M., Nikutta, R., Ivezi\'{c}, \'{Z}., \& Elitzur, M,
  2008, \apj, 685, 160
\bibitem[Onaka et al.(2007)]{IRC}   
  Onaka, T., et al. 2007, \pasj, 59, S401     
\bibitem[Piffaretti et al.(2011)]{MCXC}
  Piffaretti, R., et al. 
  % Arnaud, M., Pratt, G. W., Pointecouteau, E., \& Melin, J.-B.,
  2011, \aap, 534, 109
\bibitem[Pier \& Krolik(1993)]{smooth_torus}
  Pier, E. A., \& Krolik, J. H., 1993, \apj, 18, 673
\bibitem[Risaliti et al.(2005)]{NGC1365_1}
  Risaliti, G., et al. 
  %Elvis, M., Fabbiano, G., Baldi, A., \& Zezas, A.,
  2005, \apj, 623, L93
\bibitem[Risaliti et al.(2007)]{NGC1365_2}
  Risaliti, G., et al.
  % Elvis, M., Fabbiano, G., Baldi, A., Zezas, A.,  Salvati, M.
  2007, \apj, 659, L111
\bibitem[Risaliti et al.(2009a)]{NGC1365_3}
  Risaliti, G., et al.
  % Salvati, M., Elvis, M., Fabbiano, G., Baldi, A., Bianchi, S.,
  % Braito, V., Guainazzi, M., Matt, G., Miniutti, G.,
  % Reeves, J., Soria, R., Zezas, A.,
  2009a, \mnras, 393, L1
\bibitem[Risaliti et al.(2009b)]{NGC1365_4}
  Risaliti, G., et al.
  % Braito, V., Laparola, V., Bianchi, S., Elvis, M., Fabbiano, G.,
  % Maiolino, R., Matt, G., Reeves, J., Salvati, M., Wang, J.
  2009b, \apj, 705, L1
\bibitem[Sanders et al.(1989)]{AGN_IR}	
  Sanders, D. B.,  Phinney, E. S.,  Neugebauer, G.,
  Soifer, B. T., \& Matthews, K., 1989, \apj, 347, 29
	1989ApJ...347...29S
\bibitem[Severgnini et al.(2015)]{CT-AGN_XMM-IRAS}
  Severgnini, P., et al. 2012, \aap, 542, A46
  % Caccianiga, A.; Della Ceca, R.
\bibitem[Sugizaki et al.(2011)]{MAXI/GSC2}
  Sugizaki, M., et al. 2011, \pasj, 63, S635
\bibitem[Sunyaev \& Titarchuk(1980)]{Comptonization}
  Sunyaev, R. A., \& Titarchuk, L. G., 1980, \aap, 86, 121
\bibitem[Tachibana et al.(2016)]{CenA_MAXI}
  Tachibana, Y., et al. 
  % Kawamuro, T., Ueda, Y., Shidatsu, M., Arimoto, M.,
  % Yoshii, T., Yatsu, Y., Saito, Y., Pike, S., \& Kawai, N.,
  2016, \pasj, accepted (arXiv:1504.03208)
\bibitem[Terashima et al.(2015)]{CT-AGN_XMM-AKARI}
  Terashima, Y., et al.
  % Hirata, Yoshitaka; Awaki, Hisamitsu; Oyabu, Shinki; 
  % Gandhi, Poshak; Toba, Yoshiki; Matsuhara, Hideo
  2015, \apj, 814, 11
\bibitem[Yamamura et al.(2012)]{FIS_Catalog}
  Yamamura, I., et al. 
  % Makiuti, S., Ikeda, N., Koga, T., Yoshino, A. \& Yamauchi, C.,
  2012, Publications of The Korean Astronomical Society, 27, 105
\bibitem[Yamauchi et al.(2011)]{AKARI-CAS}
  Yamauchi, C., et al. 2011, \pasp, 123, 852
\bibitem[Ueda et al.(2007)]{new-type_AGN}
  Ueda, Y., et al. 2007, \apj, 664, L79
  % Eguchi, S., Terashima, Y., Mushotzky, R. Tueller, J.,
  % Markwardt, C. Gehrels, N., Hashimoto, Y., \& Potter, S.
\bibitem[Ueda et al.(2014)]{XRB_Ueda}
  Ueda, Y., et al. 
  % Akiyama, M., Hasinger, G., Miyaji, T. \& Watson, M.G.,
  2014, \apj, 786, 104
\bibitem[Ueda et al.(2011)]{LumFunc_MAXI}
  Ueda, Y., et al. 2011, \pasj, 63, S937  
\bibitem[Urry \& Padovani(1995)]{AGN_Unify_2}
  Urry, C. M., \& Padovani, P., 1995, \pasp, 107, 803
\end{thebibliography}
\end{document}